\documentclass[letterpaper,aps,prb,twocolumn,floatfix,amsfonts,amssymb,showpacs,superscriptaddress]{revtex4-1} 
\pdfpageattr {/Group << /S /Transparency /I true /CS /DeviceRGB>>}

\usepackage{amsmath} 
\usepackage{amsthm} 
\usepackage{amssymb} 
\usepackage{graphicx} 
\usepackage{tabularx} 
\usepackage{color} 
\usepackage{soul}
\usepackage[caption=false,justification=justified,subrefformat=parens]{subfig}

\usepackage{sidecap}

\newcommand{\comment}[1]{} 
\newcommand{\eref}[1]{Eq.~\ref{#1}}

\newcommand{\fref}[1]{Fig.~\ref{#1}} 
\newcommand{\sref}[1]{Sec.~\ref{#1}} 

\newcommand{\srefs}[2]{Secs.~\ref{#1},\ref{#2}}

\newcommand{\I}{\mathrm{i}} 
\newcommand{\E}{\mathrm{e}} 
\newcommand{\cre}[2]{#1_{#2}^\dagger} 
\newcommand{\ann}[2]{#1_{#2}^{\phantom{\dagger}}} 
\newcommand{\veccre}[2]{\vec{#1}_{#2}^{\,\,\dagger}} 
\newcommand{\vecann}[2]{\vec{#1}_{#2}^{\,\,\phantom{\dagger}}}

\newcommand{\tk}{T_\mathrm{K}} 
 
\newcommand{\boldk}{{\boldsymbol{\mathrm{k}}}} 
\newcommand{\boldq}{\boldsymbol{\mathrm{q}}} 
\newcommand{\boldr}{\boldsymbol{\mathrm{r}}}

\newcommand{\epsk}{\epsilon^{\phantom{\dagger}}_{\boldk}} 
\newcommand{\epsd}{\epsilon^{\phantom{\dagger}}_{d}} 
\newcommand{\epso}{\epsilon^{\phantom{\dagger}}_{0}}

\newcommand{\ham}{\mathcal{H}}

\newcommand{\VBZ}{\Omega_{BZ}}

\newcommand{\ev}{\mathrm{eV}} 
\newcommand{\kel}{\mathrm{K}} 
\newcommand{\sgn}[1]{\mathrm{sgn}(#1)} 
\newcommand{\creIndex}[3]{#1_{#2,\sigma}^{#3\,\,\dagger}}
\newcommand{\annIndex}[3]{#1_{#2,\sigma}^{#3\,\,\phantom{\dagger}}}

\makeatletter 
\newsavebox{\@brx} 
\newcommand{\llangle}[1][]{\savebox{\@brx}{\(\m@th{#1\langle}\)}%
  \mathopen{\copy\@brx\kern-0.5\wd\@brx\usebox{\@brx}}} 
\newcommand{\rrangle}[1][]{\savebox{\@brx}{\(\m@th{#1\rangle}\)}%
  \mathclose{\copy\@brx\kern-0.5\wd\@brx\usebox{\@brx}}} 
\makeatother

\begin{document} 
\title{Quasiparticle interference from magnetic impurities}

\author{Philip G. Derry} 
\affiliation{Oxford University, Department of Chemistry, Physical \& Theoretical Chemistry, South Parks Road, Oxford, OX1 3QZ, United Kingdom}
\author{Andrew K. Mitchell} 
\affiliation{Oxford University, Department of Chemistry, Physical \& Theoretical Chemistry, South Parks Road, Oxford, OX1 3QZ, United Kingdom}
\affiliation{Institute for Theoretical Physics, Utrecht University, Leuvenlaan 4, 3584 CE Utrecht, The Netherlands}
\author{David E. Logan} 
\affiliation{Oxford University, Department of Chemistry, Physical \& Theoretical Chemistry, South Parks Road, Oxford, OX1 3QZ, United Kingdom} 
 

\begin{abstract}
Fourier transform scanning tunneling spectroscopy (FT-STS) measures the scattering of conduction electrons from impurities and defects, giving information about the electronic structure of both the host material and adsorbed impurities. We interpret such FT-STS measurements in terms of the quasiparticle interference (QPI), here investigating in detail the QPI due to magnetic impurities adsorbed on a range of representative non-magnetic host surfaces, and  contrasting with the case of a simple scalar impurity or point defect. We demonstrate how the electronic correlations present for magnetic impurities markedly affect the QPI, showing e.g.\  a large intensity enhancement due to the Kondo effect, and universality at low temperatures/scanning-energies. The commonly-used joint density of states (JDOS) interpretation of FT-STS measurements is also considered, and shown to be insufficient in many cases, including that of magnetic impurities.
\end{abstract}

\maketitle


\section{Introduction} 
\label{sec:intro} 

Magnetic atoms embedded on non-magnetic surfaces provide realizations of quantum impurity models\cite{hewson} which are amenable to detailed experimental study and manipulation with a scanning tunneling microscope (STM).\cite{98:LiSchneiderWD,*09:TernesSchneiderWD,98:MadhavanCrommie,*01:MadhavanCrommie,00:ManoharanEigler} Such systems are promising candidates as a basis for nanoscale computational, memory storage and spintronic devices.\cite{11:HeinrichLoth,11:KhajetooriansWiebe} They are also of fundamental interest in their own right, due to the subtle interplay of strongly correlated local spin and orbital degrees of freedom coupled to a conduction electron bath. 
 
\begin{figure}
\includegraphics[width=8cm]{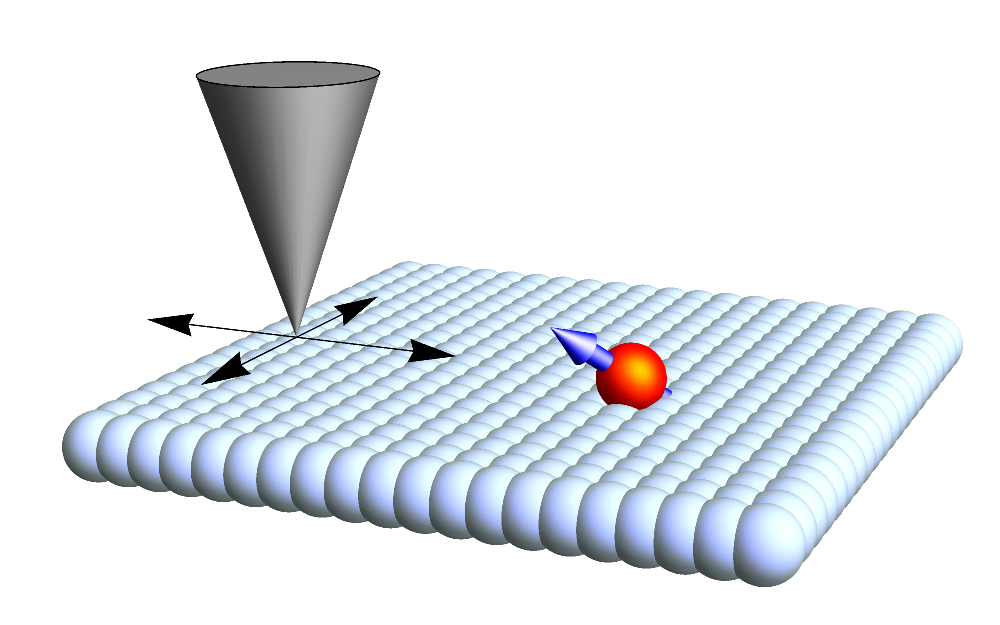}
\caption{
Schematic FT-STS setup: a spatial map of surface LDOS modulations due to electron scattering from a magnetic adatom impurity is extracted from differential conductance measurements as an STM tip is rastered over the surface.
}\label{fig:stmFig}
\end{figure}

In scanning tunneling spectroscopy (STS) experiments, the differential conductance between tip and surface is related to the local density of electronic states (LDOS) at a particular scanning energy and temperature; \cite{83:TersoffHamann,87:FeenstraStroscio} by rastering the STM tip across the surface (\fref{fig:stmFig}) a spatial map of the LDOS in the vicinity of features such as adsorbed impurities may be generated.\cite{98:MadhavanCrommie,*01:MadhavanCrommie} Such impurities break translational symmetry at the surface, causing scattering of conduction electrons and modulations in the LDOS that depend strongly on both the electronic structure of the underlying sample (`host') and the properties and distribution of impurities.  Fourier transform STS (FT-STS) -- in which such modulations are analysed in reciprocal space and interpreted in terms of the quasiparticle interference (QPI) between the diagonal states of the clean host\cite{03:WangLee,03:CapriottiScalapino,04:Markiewicz} -- thus provides a wealth of information on the nature of the impurity-host and (host-mediated) inter-impurity correlations.

The aim of this work is to calculate the QPI due to magnetic adatoms on metallic surfaces, drawing comparison with the case of non-magnetic, `scalar' impurities (s-wave potential scatterers in the weak-scattering Born limit\cite{03:CapriottiScalapino}). We also link quantitatively FT-STS measurements to the calculated QPI for generic systems. By way of contrast, we then examine the widely-used phenomenological joint density of states (JDOS) interpretation for FT-STS measurements.\cite{06:McElroyLee,11:SimonBena}

In addition to FT-STS, magnetic impurities can themselves be spectroscopically probed when the STM tip is positioned directly over an impurity. A narrow Kondo resonance is commonly observed around the Fermi energy in these local STS measurements,\cite{98:LiSchneiderWD,09:TernesSchneiderWD,98:MadhavanCrommie,*01:MadhavanCrommie} arising from the formation of a many-body Kondo singlet state where the impurity local moment is screened dynamically by surrounding host conduction electrons. 
For STS studies of Co adatoms on noble metal surfaces, typical Kondo temperatures of $\tk\sim 50-100\mathrm{K}$ are extracted from the half-width at half-maximum of the spectral Kondo resonance.\cite{04:WahlSchneider} 
The single-orbital Anderson impurity model\cite{61:Anderson} has been successfully used to rationalize such local STS measurements.\cite{00:SchillerHershfield,00:UjsaghyZawadowski,11:zitko} It is adopted here as a qualitatively accurate description of a generic, low-spin magnetic adatom; and is treated using the numerical renormalization group (NRG).\cite{nrg:rev} In other cases, e.g.\ high-spin Mn or Fe adatoms, generalized impurity models must be employed to capture the full orbital structure of the adatom and the host material\cite{FeinAu,*FeinAuII,14:WehlingWerner} (such calculations are quite specific to the particular system under consideration,\cite{13:GardonioWehling} and are not explicitly considered).

We begin by considering a general formulation 
(\srefs{sec:models}{sec:qpi}) for any number or type of impurities embedded on the surface of tight-binding hosts. The single
magnetic impurity case, which is the main focus of this paper, is simply a particular example (we subsequently apply the formalism 
to multi-impurity systems in Ref.~\onlinecite{multi_imp}).
We highlight in particular (\sref{sec:qpi})
the differences between the 3d cubic lattice with a (100) surface, the 2d square lattice, and the honeycomb lattice. These simple but representative hosts reproduce a range of possible material realizations: the LDOS for each features distinct behavior close to the Fermi level, with flat, divergent and vanishing LDOS for the three lattices respectively. Each host lattice gives rise to qualitatively different impurity physics, with e.g.\ the single-impurity Kondo temperature significantly enhanced (suppressed) by increased (depleted) density of states around the Fermi level.\cite{98:GonzalezBuxtonIngersent,13:MitchellFritzTI,13:MitchellBullaFritz,*13:MitchellFritzPLK,13:BM}

Single impurity systems are considered explicitly in \sref{sec:QPI1imp} \emph{ff}.
The QPI due to single magnetic and scalar impurities for each host is considered in \sref{sec:QPI1imp}, drawing attention to the qualitative differences in QPI arising from the different types of scattering center and host. The rich \emph{dynamical} properties of the QPI are studied in detail for systems containing a magnetic impurity in \sref{sec:characteristicKondo}; we emphasize that this strong energy-dependence cannot be reproduced in systems containing only scalar impurities or structural defects. Indeed, strong electron correlations in systems with magnetic impurities are shown to produce unique QPI signatures: in metallic systems, the QPI exhibits universality in terms of rescaled scanning energy and temperature due to the Kondo effect, while non-trivial local moment physics in observed in the 2d honeycomb case.\cite{98:GonzalezBuxtonIngersent,13:FritzVojta,00:BullaLogan,13:MitchellFritzTI,14:TuckerLogan}

\sref{sec:interp} examines the interpretation of experimentally-measurable FT-STS. By simulating the experimental protocol, we investigate the possible deviation of FT-STS measurements from calculated QPI due to the finite size of the LDOS plaquette measurements in real-space.  We conclude by critically examining the relationship between the JDOS and the QPI, showing that the two quantities are not directly related in any but the simplest case (a single scalar impurity embedded on a centrosymmetric host), where the JDOS and QPI are Hilbert conjugates.


\section{Host systems \& impurity problem} 
\label{sec:models}

\subsection{Model} 
 
We consider a host material with impurities deposited on the surface, which scatter the quasiparticles of the clean host. Initially we consider the general problem of $N$ magnetic surface adatoms, formulating the QPI generally for any type, number and distribution of adatoms.

The clean host is taken to be non-interacting and of tight-binding form, given in its real-space basis by
\begin{equation}
\label{eq:Hhost}
\ham_{\mathrm{host}} = \epso\sum_{i,\sigma}\cre{c}{\boldr_i\sigma}\ann{c}{\boldr_i\sigma}-t\sum_{\langle ij\rangle,\sigma}(\cre{c}{\boldr_i\sigma}\ann{c}{\boldr_j\sigma} + \text{H.c.}) \;,
\end{equation}
where $\cre{c}{\boldr_i\sigma}$ creates an electron of spin $\sigma=\uparrow$/$\downarrow$ in the Wannier orbital localized at site $\boldr_i$; and $\langle ij\rangle$ denotes a sum over nearest-neighbor sites, coupled by a tunnel matrix element $t$. Here we consider the half-filled host, $\epso=0$. 

Specifically, we focus on the simple 2d square lattice, the 3d cubic lattice with a (100) surface, and the 2d honeycomb lattice. The 3d cubic lattice in particular is representative of a wide class of regular metallic systems, with a constant (finite) electronic density of states at low energies. By contrast, the 2d square lattice, relevant to certain layered materials, features a van Hove singularity with a logarithmically diverging density of states around the Fermi level.\cite{13:MitchellBullaFritz,94:LudwigGrinstein} The honeycomb lattice, describing graphene within the simplest non-interacting tight-binding approximation, is notable because it is a bipartite lattice with a vanishing (pseudogapped) density of states at low energies.\cite{09:CastroNetoGeim,13:FritzVojta} Although simplified, these host systems exemplify a number of distinctive features relevant to real materials, and each induces qualitatively different impurity physics (Sec.~\ref{sec:QPI1imp}).

The full model, including $N$ impurities, is given by
\begin{equation}
\label{eq:Hfull}
\ham=\ham_{\mathrm{host}} + \sum^{N}_{\alpha=1} \ham_{\mathrm{imp},\alpha} \;.
\end{equation}

In the simplest case, the local potential at site $\boldr_{\alpha}$ is modified by impurity $\alpha$, breaking translational invariance and causing additional electronic scattering. These static defects, referred to as `potential scattering' (or `scalar') impurities, are described by
\begin{eqnarray}
\label{eq:Hps}
\ham^{\textit{ps}}_{\mathrm{imp},\alpha} = v\sum_{\sigma}\cre{c}{\boldr_\alpha\sigma}\ann{c}{\boldr_\alpha\sigma} \;. 
\end{eqnarray}
This simple model, while often appropriate to describe point defects in materials, does not faithfully capture the physics of many adsorbed impurity adatoms --- in particular magnetic impurities, which are \emph{dynamic} objects with internal degrees of freedom and strong local Coulomb interactions.\cite{hewson} These have a significant qualitative effect 
on the scattering properties, and must therefore be taken into account.
In this paper we focus primarily on adsorbed magnetic impurities, described in terms of correlated quantum levels
\begin{equation}
\label{eq:HAIM}
\begin{split}
\ham^{\textit{AIM}}_{\mathrm{imp},\alpha} =& \sum_{\sigma}\epsd\cre{d}{\alpha\sigma}\ann{d}{\alpha\sigma} + U\cre{d}{\alpha\uparrow}\ann{d}{\alpha\uparrow}\cre{d}{\alpha\downarrow}\ann{d}{\alpha\downarrow} \\
+& V\sum_{\sigma}\left( \cre{d}{\alpha\sigma}\ann{c}{\boldr_\alpha\sigma} + \text{H.c.} \right) \;, 
\end{split}
\end{equation}
where $\cre{d}{\alpha\sigma}$ creates a spin-$\sigma$ electron on impurity $\alpha$, which is coupled to host site $\boldr_\alpha$ by tunneling matrix element $V$. For simplicity we consider explicitly the particle-hole symmetric case $\epsilon_d=-U/2$. We emphasize that true magnetic impurities preserve SU$(2)$ spin symmetry and time-reversal symmetry, unlike a static polarized local moment, which simply acts like a local magnetic field.


\subsection{Impurity dynamics} 
\label{sec:impdynamics}

Single-particle dynamics of correlated impurities embedded in a non-interacting host are described generically by the Green function matrix $[\mathbf{G}_{d}(\omega)]_{\alpha,\beta}\equiv G^{\alpha\beta}_{d}(\omega) = \langle \langle \ann{d}{\alpha\sigma};\cre{d}{\beta\sigma}\rangle\rangle_{\omega}$, with $\langle \langle \hat{A};\hat{B}\rangle\rangle_{\omega}$ the Fourier transform of the retarded correlator $-i\theta(t)\langle \{ \hat{A}(t),\hat{B}(0) \}\rangle$. The matrix Dyson equation for the impurity Green functions is
\begin{equation}
\label{eq:dyson} 
\left[\mathbf{G}_{d}(\omega)\right]^{-1}=\left[\mathbf{g}_{d}(\omega)\right]^{-1}-\mathbf{\Sigma}(\omega) \;. 
\end{equation} 
The non-interacting (but host-coupled) impurity propagators are given by
\begin{equation} 
\label{eq:G0}
\left[\mathbf{g}_{d}(\omega)\right]^{-1}=(\omega+i0^{+}-\epsilon_d)\mathbf{I}-\mathbf{\Gamma}(\omega) \; ,
\end{equation} 
in terms of the hybridization matrix $\mathbf{\Gamma}(\omega)$ with elements $[\mathbf{\Gamma}(\omega)]_{\alpha,\beta} = V^2 G^0(\bold r_{\alpha},\bold r_{\beta},\omega) $; where $G^0(\bold r_{\alpha},\bold r_{\beta},\omega)=\langle \langle \ann{c}{\bold r_{\alpha}\sigma};\cre{c}{\bold r_{\beta}\sigma}\rangle\rangle^0_{\omega}$ is the propagator between sites $\bold r_{\alpha}$ and $\bold r_{\beta}$ of the clean host (without impurities), and $\bold r_{\alpha}$ and $\bold r_{\beta}$ are the host sites to which impurities $\alpha$ and $\beta$ are coupled. The self-energy matrix $\mathbf{\Sigma}(\omega)$ contains all the nontrivial information due to electronic interactions, which give rise to the Kondo effect, RKKY interaction, etc.

In the present work we employ NRG to solve the underlying quantum impurity problem.\cite{nrg:rev,wilson,*Krishnamurthy1980,*Krishnamurthy1980b} 
In the spirit of Ref.~\onlinecite{UFG}, equations of motion can be used to obtain an expression for the self-energy matrix,
\begin{equation}
\label{eq:SE}
\mathbf{\Sigma}(\omega) = [\textbf{G}_{d}(\omega)]^{-1}\textbf{F}_{d}(\omega) \;,
\end{equation}
where $[\textbf{F}_{d}(\omega)]_{\alpha,\beta}=U\langle\langle \ann{d}{\alpha\sigma};\cre{d}{\beta\sigma}\cre{d}{\beta\bar{\sigma}}\ann{d}{\beta\bar{\sigma}} \rangle\rangle_{\omega}$. Both $\textbf{G}_{d}(\omega)$ and $\textbf{F}_{d}(\omega)$ are calculated directly in NRG using the full density matrix approach\cite{fdmnrg,asbasis} within the complete Anders-Schiller basis.\cite{asbasisprl}

In the case of a \emph{single} impurity, the 
Dyson equation (Eq.~\ref{eq:dyson}) reduces to $G_{d}^{11}(\omega)=(\omega+\I 0^+ -\epsilon_d-\Gamma(\omega)  - \Sigma_{11})^{-1}$, with the hybridization 
$\Gamma(\omega)=V^2 G^0(\bold r_1,\bold r_1,\omega)$ 
related simply to the clean host LDOS (itself independent of position due to translational invariance). In NRG,\cite{nrg:rev} a discretized version of the conduction electron Hamiltonian is formulated, and mapped onto a semi-infinite 1d chain with the impurity located at one end. 
Discretizing on a logarithmic energy scale leads to the `Wilson chain' representation in which hopping matrix elements decrease exponentially down the chain. The RG scheme involves iterative diagonalization, starting at the impurity and working down the Wilson chain, discarding high-energy states at each step.\cite{nrg:rev}


\subsection{Host dynamics} 
\label{sec:hostdynamics}

The full dynamics of the host, in the presence of impurities, is embodied in the lattice Green functions $G(\bold r_{i},\bold r_{j},\omega)$ connecting arbitrary host sites $\bold r_{i}$ and $\bold r_{j}$. They can be related exactly to the above impurity Green functions by equations of motion:\cite{eom} 
\begin{align}
\label{eq:tmeqn_rs}
\begin{split}
G(\bold r_i,\bold r_j,\omega) -&G^0(\bold r_i,\bold r_j,\omega)  = \\
\sum_{\alpha,\beta} &G^0(\bold r_i,\bold r_{\alpha},\omega) T_{\alpha\beta}(\omega) G^0(\bold r_{\beta},\bold r_{j},\omega) \;,
\end{split}
\end{align}
where the sum runs over impurities $\alpha$ and $\beta$, and $T_{\alpha\beta}(\omega)$ is the real-space t-matrix. For magnetic impurities described by $\ham^{\textit{AIM}}_{\mathrm{imp}}$ in Eq.~\ref{eq:HAIM}, the t-matrix takes the form
\begin{align}
\label{eq:tm_rs}
T^{\textit{mag}}_{\alpha\beta}(\omega)=V^2G_{d}^{\alpha\beta}(\omega) \;,
\end{align}
requiring as such a knowledge of the full impurity Green functions from Eq.~\ref{eq:dyson}. By contrast, the t-matrix for potential scattering impurities, described by $\ham^{\textit{ps}}_{\mathrm{imp}}$ in Eq.~\ref{eq:Hps}, can be obtained simply in closed form,
\begin{align}
\label{eq:tmps_rs}
T^{\textit{ps}}_{\alpha\beta}(\omega)=v[\textbf{I}-v\textbf{G}^0(\omega)]^{-1}_{\alpha\beta} \;,
\end{align}
where the elements $[\textbf{G}^0(\omega)]_{\alpha'\beta'}\equiv G^0(\bold r_{\alpha'},\bold r_{\beta'},\omega)$ are free host Green functions.

In the diagonal quasiparticle basis of the clean host, 
\begin{equation}
\label{eq:Hhostk}
\ham_{\mathrm{host}} = \sum_{\bold k,\sigma}\epsilon^{\phantom{\dagger}}_{\bold k} \cre{c}{\bold k \sigma}\ann{c}{\bold k\sigma} \;
\end{equation}
with $\epsilon_{\bold k}$ the dispersion ($\bold k$ labels the Bloch state momentum). The t-matrix equation (Eq.~\ref{eq:tmeqn_rs}) can also be transformed into the momentum-space basis,
\begin{equation} 
\label{eq:tmeqn_k}
\Delta G(\boldk,\boldk',\omega) = G^{0}(\boldk,\omega)T(\boldk,\boldk',\omega)G^{0}(\boldk',\omega) \;, 
\end{equation} 
where $\Delta G(\boldk,\boldk',\omega)=G(\boldk,\boldk',\omega)-G^{0}(\boldk,\omega)\delta_{\bold k,\bold k'}$ and $G^{0}(\boldk,\omega)=(\omega+\I0^+-\epsilon_\boldk)^{-1}$. 
All quasiparticle scattering induced by the impurities is now contained in 
\begin{equation} 
\label{eq:tm_k}
T(\boldk,\boldk',\omega)= \frac{1}{\VBZ}\sum_{\alpha,\beta}\text{e}^{\I(\bold k' \cdot \bold r_{\beta}- \bold k \cdot \bold r_{\alpha})} \times T_{\alpha\beta}(\omega)\;,
\end{equation} 
where $\VBZ$ is the volume of the first Brillouin zone.


\section{QPI} 
\label{sec:qpi}

At sufficiently low temperature and bias, the differential conductance $dI(\boldr,V)/dV$ between STM tip and surface for a particular bias $V$ is 
proportional to the LDOS, $\rho(\boldr,\omega=\mathrm{e}V)$, of the sample at position $\boldr$. STS thus measures the energy-resolved electronic structure of the sample at a particular point in real space.\cite{83:TersoffHamann,87:FeenstraStroscio}

The experimental quantity of interest is the QPI,
\begin{align}\label{eq:rhoDFTdeft}
\rho(\boldq,\omega)=\sum_{i\in(L \times L)}\E^{-\I\boldq\cdot\boldr_i}\rho(\boldr_i,\omega) \;,
\end{align}
obtained at given scanning energy $\omega$. It is therefore the Fourier transform of a real-space LDOS map $\rho(\bold r_i, \omega)$, itself measured by STS over an $L\times L$ sample region.
Since the STS experiment probes the \emph{surface}, the LDOS map is two-dimensional. In 2d systems such as graphene, or effective 2d layered systems such as the cuprates, STS thus probes directly the underlying lattice. In 3d systems, by contrast, the lattice probed by STS corresponds to the crystallographic surface lattice.

The FT-STS technique has been employed to investigate the electronic structure of materials and to map their Fermi surface contours,\cite{98:PetersenHoffman,*00:PetersenHoffman,11:SimonBena} providing complementary information to techniques such as ARPES.\cite{04:Markiewicz, 06:McElroyLee} FT-STS has proven of particular value in the study of layered materials such as cuprates and pnictides,\cite{02:HoffmanMcElroy,03:HowaldKapitulnik,*07:FischerRenner,09:YinHoffman,*09Rev:YinHoffman,*09:ParkHinkov} as well as topological insulators,\cite{09:RoushanYazdani,*09:ZhangCheng} graphene\cite{08:BrihuegaVeuillen,*09:ZhangCrommie} and heavy fermion materials.\cite{13:ToldinKirchnerMorr,*12:TingFigginsMorr}
In many cases (e.g.\ for weak, disorder-induced scattering in cuprates\cite{02:HoffmanMcElroy}), the local defects giving rise to the QPI may be approximated as scalar impurities in the Born limit;\cite{03:CapriottiScalapino} 
although even for scalar impurities this simplification is known to be insufficient in some cases.\cite{13:ToldinKirchnerMorr,*12:TingFigginsMorr}
For e.g.\ transition metal adatoms, however, the full interacting impurity model must be considered. 

The QPI at $\boldq=\bold 0$ is often omitted in experimental results, because it corresponds to the \emph{total} density of states sampled, and is extensive in $L^2$. 
The desired impurity contribution to the QPI is then obtained by subtracting the result for the clean system without impurities, 
\begin{eqnarray}\label{eq:QPI_def} 
\Delta\rho(\boldq,\omega)=\sum_{i\in(L \times L)}\E^{-\I\boldq\cdot\boldr_i}\Delta\rho(\boldr_i,\omega) \; 
\end{eqnarray}
where $\Delta\rho=\rho-\rho^0$ (with $\rho^0$ for the clean host). Since $\rho^0(\boldq,\omega)\propto \delta_{\bold q,\bold 0}$, $\Delta\rho(\boldq,\omega)$ scales with the number of impurities. The normalized QPI power spectrum, $|\Delta\rho(\boldq,\omega)/N|^2 \propto |\Delta\rho(\boldq,\omega)/L^2|^2 $, is \emph{intensive}, independent of the number of impurities or sample region size.

We now give a general formulation for calculating the QPI due to scattering from single or multiple impurities, which can either be simple static potential scattering defects or magnetic (dynamic, interacting) impurities.


\subsection{Real-space formulation} 
\label{sec:qpi_rs}
 Following the experimental protocol, the QPI can be calculated by discrete Fourier transform of the LDOS within an $L\times L$ region of the host surface, using Eq.~\ref{eq:QPI_def}. The LDOS at site $\bold r_i$ in the presence of impurities is related to the local host Green function, $\rho(\bold r_i,\omega)=-\tfrac{1}{\pi}\text{Im}~G(\bold r_i, \bold r_i,\omega)$, such that from Eq.~\ref{eq:tmeqn_rs},
\begin{equation}
\label{eq:delta_rho}
\Delta \rho(\bold r_i,\omega) = -\frac{1}{\pi}\text{Im}~\sum_{\alpha,\beta} G^0(\bold r_i,\bold r_{\alpha},\omega) T_{\alpha\beta}(\omega) G^0(\bold r_{\beta},\bold r_{i},\omega) \;
\end{equation}
in terms of the full scattering t-matrix and free non-local host Green functions. The virtue of Eq.~\ref{eq:delta_rho} is that it is entirely general, and can in principle be used for any lattice with any number or type of impurities. Although the LDOS is \emph{sampled} over a finite region, we stress that the host system is in the thermodynamic limit. 

The accurate calculation of lattice Green functions $G^0(\bold r_i, \bold r_j,\omega)$ is itself a subtle and well-studied problem.\cite{velev,71:KatsuraMoritaINTRO,10:Guttmann,00:Cserti} Exact diagonalization of finite-sized lattices or discrete Fourier transforms yield poor approximations to Green functions of the desired (semi-)infinite systems, especially at low scanning energies or near van Hove singularities. Recursion methods\cite{71:KatsuraMoritaINTRO,71:MoritaSC,*71:MoritaRECUR} can be used for periodic systems where the exact dispersion $\epsilon_{\bold k}$ is known --- but such techniques are numerically unstable for large site separations $|\bold r_i-\bold r_j|$, and converged solutions can be computationally demanding.\cite{09:Berciu,*10:BerciuCook} Although `bond cutting'\cite{pollman} and `continued fraction'\cite{72:HaydockKelly,*75:HaydockKelly,*75:FalicovYndurain,85:PettiforWeaire} variants have been developed in special cases, recursion methods typically cannot be used for systems with a surface that breaks translational invariance, limiting applicability within the QPI context. In consequence, calculation of all $N\times L^2$ non-local Green functions required for a system of $N$ impurities in an $L\times L$ region is often the major challenge.

In this paper we have developed instead a novel technique for fast and accurate numerical calculation of free Green functions on hypercubic-type lattices. The method is detailed in the appendix, and involves successive convolutions of simpler 1d Green functions which are known exactly in closed form. The convolution itself can moreover be performed efficiently using fast Fourier transform.


\subsection{Scattering state formulation} 
\label{sec:qpi_tm}
The `true' QPI containing all scattering information is obtained by taking the thermodynamic limit of plaquette size, $L\rightarrow\infty$. Translational symmetry implies a basis of states with well defined momentum \emph{parallel to the surface}, i.e. over the first \emph{surface} Brillouin zone (1SBZ). The local Green function for a surface lattice site is expressed in terms of this basis by 2d Fourier transformation, 
\begin{equation}
\label{eq:Gkpar}
G(\boldr_i,\boldr_i,\omega)= \iint\limits_{1SBZ}\frac{d^2\mathrm{\bold{k_{\parallel}}}d^2\mathrm{\bold{k_{\parallel}'}}}{\VBZ}~\E^{-\I\boldr_i\cdot(\bold{k_{\parallel}'}-\bold{k_{\parallel}})}\times G(\bold{k_{\parallel}},\bold{k_{\parallel}'},\omega) \;, 
\end{equation}
with $\VBZ$ the volume of the 1SBZ. Writing $\Delta\rho(\bold r_i,\omega)=-\tfrac{1}{\pi}\text{Im}~\Delta G(\bold r_i,\bold r_i,\omega)$, Eq.~\ref{eq:QPI_def} takes the form,
\begin{align}
\label{eq:QPI_k} 
\begin{split}
\Delta\rho(\boldq,\omega)=&-\frac{1}{\pi}\sum_i\E^{-\I\boldq\cdot\boldr_i}  \\  
&\times\text{Im} \iint\limits_{1SBZ}\frac{d^2\mathrm{\bold{k_{\parallel}}}d^2\mathrm{\bold{k_{\parallel}'}}}{\VBZ}~\E^{-\I\boldr_i\cdot(\bold{k_{\parallel}'}-\bold{k_{\parallel}})} \Delta G(\bold{k_{\parallel}},\bold{k_{\parallel}'},\omega) \;,
\end{split}
\end{align}
where $\Delta G(\bold{k_{\parallel}},\bold{k_{\parallel}'},\omega)= G(\bold{k_{\parallel}},\bold{k_{\parallel}'},\omega)-G^0(\bold{k_{\parallel}},\bold{k_{\parallel}'},\omega)$. Eq.~\ref{eq:QPI_k} can itself be recast as
\begin{eqnarray}
\label{eq:QPI_Q} 
\Delta\rho(\boldq,\omega)=-\frac{1}{2\pi \I} \left [ Q(\bold q,\omega) - Q(-\bold q, \omega)^* \right ] \;,
\end{eqnarray}
where 
\begin{subequations}
\label{eq:Q_def} 
\begin{align}
Q(\bold q,\omega) &= \int\limits_{1SBZ}d^2\mathrm{\bold{k_{\parallel}}}~\Delta G(\bold{k_{\parallel}},\bold{k_{\parallel}-\bold q},\omega) \; \\
& \equiv \sum_{\alpha,\beta}T_{\alpha\beta}(\omega)\times \Lambda_{\alpha\beta}(\bold q,\omega) \;.
\end{align}
\end{subequations}
As highlighted by Eq.\ \ref{eq:Q_def}b, the QPI factorises into a momentum-independent scattering amplitude $T_{\alpha\beta}(\omega)$ (Eqs.\ \ref{eq:tm_rs} or \ref{eq:tmps_rs}), and a host response function $\Lambda_{\alpha\beta}(\bold q,\omega)$ which depends only on the host lattice and the spatial location of impurities, but not the type of impurity (and thus details of the scattering). The explicit form of this host function must be determined separately for each lattice, as considered below.


\subsubsection{2d square lattice} 
\label{sec:qpi_2d}
Consider first the 2d square lattice, where the QPI calculation is simplest. As the system is itself two-dimensional, the surface-momentum basis is simply the diagonal representation,  $\bold k = \bold k_{\parallel}$. The QPI thus follows from Eqs.\ \ref{eq:Q_def}a,\ref{eq:tmeqn_k},\ref{eq:tm_k}, and is indeed of form Eq.\ \ref{eq:Q_def}b with
\begin{equation}
\label{eq:lambda_def_sq} 
\Lambda_{\alpha\beta}(\bold q,\omega) = \int\limits_{1SBZ}\frac{d^2\mathrm{\bold{k}}}{\VBZ}~G^{0}(\boldk,\omega)G^{0}(\boldk-\bold q,\omega) \E^{\I[\bold k \cdot \bold r_{\alpha}- (\bold k-\bold q) \cdot \bold r_{\beta}]} \;.
\end{equation}
The free momentum-space Green functions are themselves given by\cite{note:eta} $G^{0}(\boldk,\omega)=(\omega+\I0^+-\epsilon_{\boldk})^{-1}$, with 2d square lattice dispersion (and lattice  constant $\mathrm{a}_0$)
\begin{equation}
\label{eq:2dsq_disp}
\epsilon_{\boldk}=-2t[\cos(\mathrm{a}_0 \mathrm{k}_x)+\cos(\mathrm{a}_0 \mathrm{k}_y)] \;.
\end{equation}
The half-bandwidth is then $D=4t$ in terms of the lattice hopping matrix element $t$ appearing in Eq.~\ref{eq:Hhost}.

$\Lambda_{\alpha\beta}(\bold q,\omega)$ can be computed efficiently by using the convolution theorem to do the Brillouin zone integration:
\begin{align}
\mathcal{F}_{\bold k}[\Lambda_{\alpha\beta}(\bold q,\omega)] = \mathcal{F}_{\bold k}[G^{0}(\boldk,\omega)\E^{\I\bold k \cdot \bold r_{\alpha}}] \times \mathcal{F}_{\bold k}[G^{0}(\boldk,\omega) \E^{-\I\bold k \cdot \bold r_{\beta}}] 
\nonumber
\end{align}
where $\mathcal{F}_{\bold k}$ denotes the 2d fast Fourier transform.


\subsubsection{3d cubic lattice with (100) surface} 
\label{sec:qpi_3dcubic}
3d host lattices are more subtle, due to the surface-sensitive STM measurement. As only the surface LDOS is probed, the QPI amounts to a partial trace over the t-matrix equation Eq.~\ref{eq:tmeqn_k}, in contrast to the full trace for the 2d square lattice (Eq.~\ref{eq:lambda_def_sq}). The QPI must thus be evaluated in a basis which preserves the layer index, the surface momentum basis of Eq.~\ref{eq:Gkpar} (rather than the diagonal basis of Eq.~\ref{eq:tmeqn_k}). $\Delta G(\bold{k_{\parallel}},\bold{k_{\parallel}-\bold q},\omega)$ in Eq.~\ref{eq:Q_def} thus involves propagators between states with surface momentum $\bold{k_{\parallel}}$ and $\bold{k_{\parallel} -q}$. The 2d transform of $G(\boldr_i,\boldr_i,\omega)$, Eq.~\ref{eq:Gkpar}, leads to a diagonal representation in each 2d plane in isolation -- but surface states labelled by $\bold k_{\parallel}$ remain
coupled to the bulk (and thus to each other). In general, $\Delta G(\bold{k_{\parallel}},\bold{k_{\parallel}-\bold q},\omega)$ does not therefore take the form of Eq.~\ref{eq:tmeqn_k}, but rather
\begin{align}
\label{eq:tmeqn_gen}
\begin{split}
&\Delta G(\boldk_{\parallel},\boldk'_{\parallel},\omega) = \\& \iint\limits_{1SBZ}\frac{d^2\mathrm{\bold{k_{\parallel}''}}d^2\mathrm{\bold{k_{\parallel}'''}}}{\VBZ}~\tilde{G}^{0}(\boldk_{\parallel},\boldk_{\parallel}'',\omega)T(\boldk_{\parallel}'',\boldk'''_{\parallel},\omega)\tilde{G}^{0}(\boldk'''_{\parallel},\boldk'_{\parallel},\omega) \;, 
\end{split}
\end{align}
where  $\tilde{G}^{0}(\boldk_{\parallel},\boldk_{\parallel}'',\omega)$ is a complex through-bulk propagator. Calculation of the QPI thus in general requires the full integrals over intermediate scattering pathways.\\

In the case of hypercubic-type lattices, however, a significant simplification arises, because $\boldk_{\parallel}$ is still a good quantum number; so a surface momentum t-matrix equation with the same structure as Eq.~\ref{eq:tmeqn_k} still applies, albeit with modified host surface Green functions. The result for the 3d cubic lattice with a (100) surface is simply,
\begin{equation}
\label{eq:tmeqn_kpar}
\Delta G(\boldk_{\parallel},\boldk'_{\parallel},\omega) = G_{\text{surf}}^{0}(\boldk_{\parallel},\omega)T(\boldk_{\parallel},\boldk'_{\parallel},\omega)G_{\text{surf}}^{0}(\boldk'_{\parallel},\omega) \;, 
\end{equation}
where the t-matrix is still given by Eq.~\ref{eq:tm_k}, but 
\begin{align}
\label{eq:G01d}
\begin{split}
G_{\text{surf}}^{0}(\boldk_{\parallel},\omega) = f\left( \frac{\omega-\epsilon_{\bold k_{\parallel}}}{2t}\right ) 
\qquad \text{where}\\
tf(\tilde{\omega}) = \tilde{\omega} -\begin{cases} \sgn{\tilde{\omega}}\sqrt{{\tilde{\omega}}^2-1}\quad &|\tilde{\omega}|>1\\\I\sqrt{1-\tilde{\omega}^2} \quad &|\tilde{\omega}|\leq1\end{cases} 
\end{split} 
\end{align}
with $\epsilon_{\bold k_{\parallel}}$ the \emph{2d square lattice} dispersion, Eq.~\ref{eq:2dsq_disp}. This broadens the pole in $G^0(\bold k,\omega)$ arising for the pure 2d system, to an ellipse of width $2t$ centered on $\omega=\epsilon_{\bold k_{\parallel}}$ in $G_{\text{surf}}^{0}(\boldk_{\parallel},\omega)$ for the 3d system. It follows that the structure of the host function $\Lambda_{\alpha\beta}(\boldq,\omega)$ is the same as in the 2d square case, Eq.~\ref{eq:lambda_def_sq}, with $G^0_\text{surf}(\bold{k_{||}},\omega)$ in place of $G^0(\boldk,\omega)$. In this paper we calculate the true 3d cubic QPI via Eqs.~\ref{eq:QPI_Q}, \ref{eq:Q_def}, \ref{eq:lambda_def_sq}, using the exact expression for the bulk-coupled surface Green functions, Eq.~\ref{eq:G01d}. 

For more complex systems where such a prescription is not available, the bulk dephasing of pure 2d surface states could be approximated by using  $G^{0}(\boldk_{\parallel},\omega)=(\omega+\I\eta-\epsilon_{\bold k_{\parallel}})^{-1}$, with finite $\eta>0$. Green function poles are thereby lifetime broadened by lorentzians of width $\eta$.


\subsubsection{Honeycomb lattice} 
\label{sec:qpi_hc}
The honeycomb lattice is complicated by the bipartite nature of the lattice, which is viewed as two interlocking triangular sublattices.  We define
\begin{equation}
\label{eq:hc_t}
t(\bold{k})=\E^{-\I \bold{k}\cdot\boldsymbol\tau}\left [ 1+\E^{\I \bold k \cdot \bold a_1}+\E^{\I \bold k \cdot \bold a_2} \right ] 
\end{equation}
in terms of the triangular sublattice vectors $\bold a_1$ and $\bold a_2$ and the inter-sublattice vector $\boldsymbol \tau=\bold r_i^A-\bold r_i^B$ (where $\bold r_i^{\gamma}$ is a site $i$ on the $\gamma=A/B$ sublattice). The $A/B$-sublattice structure gives rise to distinct $+/-$ bands in momentum space.\cite{09:CastroNetoGeim} The honeycomb lattice dispersion for these $+/-$ bands, and the complex phase, follow as
\begin{subequations}
\begin{align}
\label{eq:hc_disp}
\epsilon_{\bold k}^{\pm} &= \pm \left |t(\bold k)\right | \\
\label{eq:hc_phase}
\phi(\bold k) &= \arg\left [t(\bold k)\right ] \;.
\end{align}
\end{subequations}
Real-space operators are expressed in a diagonal basis by,
\begin{equation}
\label{eq:hc_op_trans}
\ann{c}{\bold r_i^{\gamma}} = \frac{1}{\sqrt{2}} \int\limits_{1BZ}\frac{d^2\mathrm{\bold{k}}}{\VBZ^{1/2}}~ 
\E^{\I\bold r_i^{\gamma}\cdot \bold k} ~\E^{\I s_{\gamma} \phi(\bold k)/2} \left [ \ann{c}{-,\bold k} + s_{\gamma} \ann{c}{+,\bold k} \right ] \;,
\end{equation}
where $s_{\gamma}=\pm 1$ for the $A$ or $B$ sublattice.

Generalizing Eqs.~\ref{eq:QPI_def} and \ref{eq:delta_rho} to take account of this sublattice structure gives 
\begin{equation}
\label{eq:hc_qpi_sum}
\Delta \rho(\bold q,\omega) = \sum_{\gamma,\gamma_1,\gamma_2} \Delta \rho_{\gamma,\gamma_1,\gamma_2}(\bold q, \omega) \;,
\end{equation}
where
\begin{align}
\label{eq:hc_qpi_decomp}
\begin{split}
\Delta &\rho_{\gamma,\gamma_1,\gamma_2}(\bold q, \omega) = -\frac{1}{\pi} \sum_{i\in \gamma} \E^{-\I \bold q \cdot \bold r_i^{\gamma}}  \\
&\times\text{Im}\sum_{\alpha \in \gamma_1} \sum_{\beta \in \gamma_2} G^0(\bold r_i^{\gamma},\bold r_{\alpha}^{\gamma_1},\omega) T_{\alpha\beta}(\omega) G^0(\bold r_{\beta}^{\gamma_2},\bold r_{i}^{\gamma},\omega) 
\end{split}
\end{align}
with the real-space sum over $i$ spanning sites $\bold r_{i}^{\gamma}$ on sublattice $\gamma$. Impurity $\alpha$($\beta$) is taken to be on sublattice $\gamma_1$($\gamma_2$). Thus, $\Delta \rho_{\gamma,\gamma_1,\gamma_2}(\bold q, \omega)$ is the contribution to the full QPI from sites on the $\gamma$ sublattice due to impurity-induced scattering between $\gamma_1$ and $\gamma_2$ sublattices.

Using Eq.~\ref{eq:hc_op_trans} in the definition $G^0(\bold r_i^{\gamma},\bold r_{j}^{\gamma'},\omega)=\langle \langle \ann{c}{\bold r_i^{\gamma}} ; \cre{c}{\bold r_j^{\gamma'}} \rangle \rangle_{\omega}^0$, $\Delta \rho_{\gamma,\gamma_1,\gamma_2}(\bold q, \omega)$ takes the same form as Eq.~\ref{eq:QPI_k},
\begin{align}
\label{eq:hc_QPI_k} 
\begin{split}
\Delta\rho&_{\gamma,\gamma_1,\gamma_2}(\boldq,\omega)=-\frac{1}{\pi}\sum_i\E^{-\I\boldq\cdot\boldr_i^{\gamma}} \\  
&\times \text{Im} \iint\limits_{1BZ}\frac{d^2\mathrm{\bold{k}}d^2\mathrm{\bold{k'}}}{\VBZ}~\E^{-\I\boldr_i^{\gamma}\cdot(\bold{k'}-\bold{k})} \Delta G_{\gamma,\gamma_1,\gamma_2}(\bold{k},\bold{k'},\omega) \;,
\end{split}
\end{align}
but now with
\begin{align}
\label{eq:hc_DeltaG} 
\begin{split}
\Delta G&_{\gamma,\gamma_1,\gamma_2}(\bold{k},\bold{k'},\omega) = \sum_{\alpha \in \gamma_1} \sum_{\beta \in \gamma_2}\frac{T_{\alpha\beta}(\omega)}{4\VBZ}\\
&\times \E^{-\I(\boldr_{\alpha}^{\gamma_1}\cdot\bold{k}-\bold r_{\beta}^{\gamma_2}\cdot \bold k')} ~\E^{\I[(s_{\gamma}-s_{\gamma_1})\phi(\bold k)-(s_{\gamma}-s_{\gamma_2})\phi(\bold k')]/2} \\
&\times \left (s_{\gamma} G^0_{-}(\bold k) + s_{\gamma_1} G^0_{+}(\bold k)\right )\left (s_{\gamma} G^0_{-}(\bold k') + s_{\gamma_2} G^0_{+}(\bold k')\right) \;,
\end{split}
\end{align}
in terms of the $+/-$ band free Green functions\cite{note:eta} $G_{\pm}^0(\bold k)\equiv \langle \langle \ann{c}{\pm,\bold k}; \cre{c}{\pm,\bold k}\rangle \rangle_{\omega}^0 = (\omega + \I 0^+ - \epsilon_{\bold k}^{\pm})^{-1}$. The QPI contribution then follows as,
\begin{equation}
\label{eq:hc_QPI_Q} 
\Delta\rho_{\gamma,\gamma_1,\gamma_2}(\boldq,\omega)=-\frac{1}{2\pi \I} \left [ Q_{\gamma,\gamma_1,\gamma_2}(\bold q,\omega) - Q_{\gamma,\gamma_1,\gamma_2}(-\bold q, \omega)^* \right ] 
\end{equation}
where 
\begin{eqnarray}
\label{eq:hc_Q_def} 
Q_{\gamma,\gamma_1,\gamma_2}(\bold q,\omega) = \sum_{\alpha \in \gamma_1} \sum_{\beta \in \gamma_2}T_{\alpha\beta}(\omega)\times \Lambda^{\gamma,\gamma_1,\gamma_2}_{\alpha\beta}(\bold q,\omega) .~~
\end{eqnarray}
Eqs.~\ref{eq:hc_QPI_Q} and \ref{eq:hc_Q_def} are thus analogues of Eqs.~\ref{eq:QPI_Q} and \ref{eq:Q_def}, with 
\begin{equation}
\label{eq:hc_lambda_def} 
\begin{split}
&\Lambda^{\gamma,\gamma_1,\gamma_2}_{\alpha\beta}(\bold q,\omega) =\\
& \int\limits_{1BZ}\frac{d^2\mathrm{\bold{k}}}{4\VBZ} \E^{-\I(\boldr_{\alpha}^{\gamma_1}\cdot\bold{k}-\bold r_{\beta}^{\gamma_2}\cdot (\bold k - \bold q))} \E^{\I[(s_{\gamma}-s_{\gamma_1})\phi(\bold k)-(s_{\gamma}-s_{\gamma_2})\phi(\bold k-\bold q)]/2} \\
&\times \left (s_{\gamma} G^0_{-}(\bold k) + s_{\gamma_1} G^0_{+}(\bold k)\right )\left (s_{\gamma} G^0_{-}(\bold k-\bold q) + s_{\gamma_2} G^0_{+}(\bold k - \bold q)\right) 
\end{split}
\end{equation}
We stress that interband scattering and the momentum-dependent phase factors appearing in Eq.~\ref{eq:hc_lambda_def} are important, and affect the full QPI qualitatively.


\section{Single impurity QPI:\\ magnetic \& scalar impurities} 
\label{sec:QPI1imp} 

The generalized problem involving $N$ magnetic impurities, spatially separated and coupled to conduction electrons of the host lattice, is naturally highly rich and complex (we consider aspects of it in subsequent  work\cite{multi_imp}). From here on in this paper, we focus on a single magnetic impurity -- in terms of which experimental QPI patterns are in fact typically interpreted. In this case the QPI is given by Eq.~\ref{eq:Q_def}b (with $\alpha =\beta =1$, dropped hereafter),
\begin{equation}
\label{eq:Q_1imp}
Q(\bold q,\omega) = T(\omega)\Lambda(\bold q,\omega) \;,
\end{equation}
with $T(\omega)$ the single-impurity t-matrix and explicit forms for the host functions $\Lambda(\bold q,\omega)$ given in Secs.~\ref{sec:qpi_1imp_2d}-\ref{sec:qpi_1imp_hc}.

The impurity itself is often taken to be a static potential defect in the weak-scattering Born limit.\cite{03:CapriottiScalapino} The t-matrix is then pure real and energy-independent,
\begin{equation}
\label{eq:born}
T^{\textit{ps}}(\omega)\simeq v \;,
\end{equation}
with $v$ the potential scattering strength (see Eq.~\ref{eq:Hps}). Eq.~\ref{eq:born} is the leading-order approximation to the exact Eq.~\ref{eq:tmps_rs}, holding provided $|v G^0(\bold r_{\alpha},\bold r_{\alpha},\omega)|\ll 1$. We add however that this approximation is \emph{not} valid in the vicinity of divergences in the host density of states (arising e.g.\ at $\omega=0$ in the 2d square lattice). In the special case of a single impurity on a centrosymmetric surface, $Q(\bold q,\omega)=Q(-\bold q,\omega)$, so (Eq.~\ref{eq:QPI_Q}) $\Delta \rho(\bold q, \omega) =-\frac{1}{\pi}\text{Im}~Q(\bold q,\omega)$. For a scalar impurity in the Born limit,
\begin{equation}
\label{eq:qpi_1imp_cs_ps}
\Delta \rho(\bold q, \omega) ~\overset{\text{scalar}}{=}~-\frac{v}{\pi}\Lambda''(\bold q,\omega) 
\end{equation}
where $\Lambda(\bold q,\omega)=\Lambda'(\bold q,\omega)+\I\Lambda''(\bold q,\omega)$. The QPI scanning-energy dependence is thus due solely to that of $\Lambda''(\bold q,\omega)$.

\begin{figure*}[t]
\includegraphics[width=16cm]{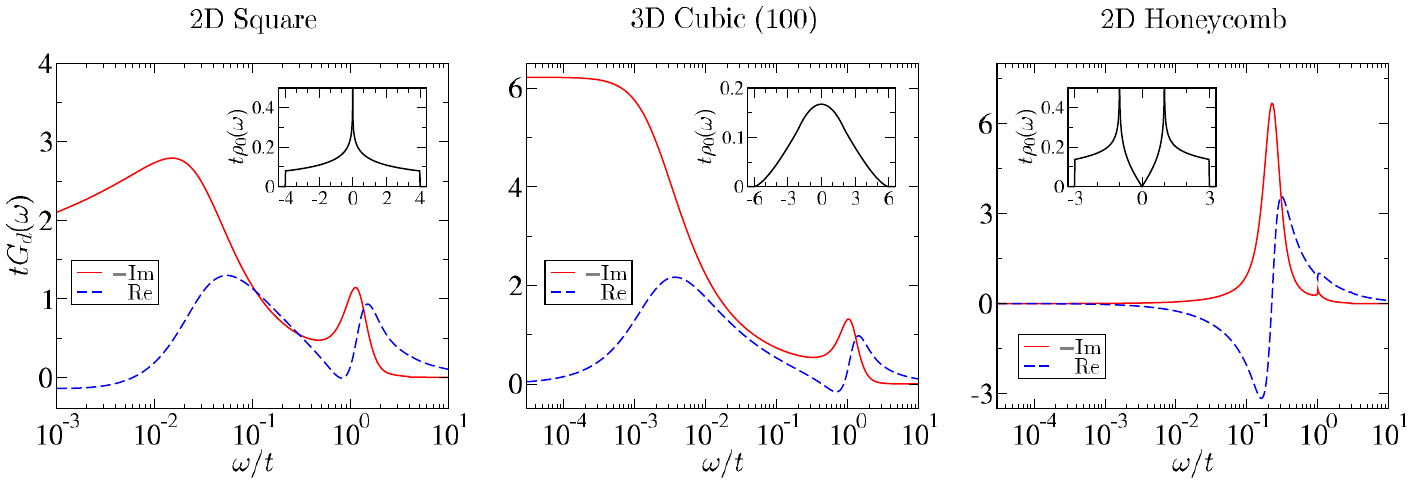}
\caption{
Dynamics of a single magnetic impurity on various lattices at $T=0$, plotted as $-\text{Im}~tG_{d}(\omega)$ (solid lines) and $\text{Re}~tG_{d}(\omega)$ (dashed) vs $\omega/t$, calculated via NRG. Insets show corresponding host density of states $t\rho_0(\omega)$. Impurity parameters: $U=1.95t$ and $V=0.555t$ for 2d square and 3d cubic lattices; $U=0.704t$ and $V=0.493t$ for honeycomb lattice. 
}\label{fig:Gimp}
\end{figure*}

For magnetic impurities by contrast, electron correlations give rise to nontrivial dynamics. From Eq.~\ref{eq:tm_rs}, 
\begin{equation}
\label{eq:Tsiam}
T^{\textit{mag}}(\omega)= V^2G_{d}(\omega) 
\end{equation}
in terms of the impurity Green function $G_{d}(\omega)$; the QPI follows from Eqs.\ref{eq:QPI_Q},\ref{eq:Q_1imp}. For a centrosymmetric surface,
\begin{equation}
\label{eq:qpi_1imp_cs_mag}
\begin{split}
\Delta \rho(\bold q, \omega) \overset{\text{mag}}{=} -\frac{V^2}{\pi} \big [ \text{Re}~&G_{d}(\omega)\Lambda''(\bold q,\omega) \\ &+ \text{Im}~G_{d}(\omega)\Lambda'(\bold q,\omega) \big ] \;,
\end{split}
\end{equation}
with contributions from both real and imaginary parts of $\Lambda(\bold q,\omega)$, and weights that depend on the impurity Green function at energy $\omega$. As discussed below, the Kondo effect produces a scattering enhancement at low temperatures and scanning energies, causing a crossover in the QPI from being dominated by $\Lambda''(\bold q,\omega)$ at high energies (similar to that of a scalar impurity) to being dominated by $\Lambda'(\bold q,\omega)$ at low energies.


\subsection{Effect of host on impurity dynamics}
For a magnetic impurity, the QPI depends on both the host function $\Lambda(\bold q,\omega)$ and the impurity Green function $G_{d}(\omega)$ -- which itself depends on the host. Specifically, the impurity problem is controlled by the hybridization function $\Gamma(\omega)$, related to the clean host density of states, $\rho_0(\omega)$, by $-\text{Im}~\Gamma(\omega)=\pi V^2 \rho_0(\omega)$. The Kondo physics is sensitive to the behavior of $\rho_0(\omega)$ near the Fermi level 
($\omega =0$);\cite{98:GonzalezBuxtonIngersent,13:MitchellFritzTI,13:MitchellBullaFritz,13:MitchellFritzPLK,14:TuckerLogan} and to leading order,
\begin{equation}
\label{eq:dos}
\rho_0(\omega) \overset{|\omega|\ll t}{\sim} 
\begin{cases}
\frac{\log(16 t/|\omega|)}{2\pi^2t} \qquad &:~\text{2d square} \\
\tfrac{1}{6t}-\tfrac{\omega^2}{6\sqrt{2}\pi^2t^3} \qquad &:~\text{3d cubic, (100) surface} \\
\frac{|\omega|}{\sqrt{3}\pi t^2} \qquad &:~\text{honeycomb} \;.
\end{cases}
\end{equation}
These lattices exemplify three paradigms, with densities of states that are diverging, flat, or pseudogapped at low-energy. $\rho_0(\omega)$ \emph{vs} $\omega/t$ is shown in the insets to Fig.~\ref{fig:Gimp}.

The density of states for metallic systems is typically flat at low energies. This gives rise to an exponentially-small Kondo scale\cite{hewson} 
$T_K/t\sim \exp[-\pi U/8V^{2}\rho_{0}(0)]$, and low-energy Fermi liquid behavior\cite{hewson}
\begin{equation}
\label{eq:specpin}
\text{Im}\Gamma(\omega) \times\text{Im}G_{d}(\omega) 
~\overset{|\omega|\ll T_K}{\sim}~ 1 -\alpha_{\omega}(\omega/T_K)^2+... \;,
\end{equation}
with Fermi level spectral pinning, $-\pi V^{2}\rho_{0}(0)\mathrm{Im}G_{d}(0)=1$ (as $\text{Im}\Gamma(0)=-\pi V^{2}\rho_{0}(0)$). This is shown for a magnetic impurity on the (100) surface of a 3d cubic lattice in Fig.~\ref{fig:Gimp} (center panel), where the imaginary and real parts of $tG_{d}(\omega)$ are plotted vs $\omega/t$. We have chosen representative impurity parameters $U=1.95t$ and $V=0.56t$, yielding $T_K\approx 5\times 10^{-3}t$ [defined here as the half-width at half-maximum of the Kondo resonance\cite{note:tkdef}]. With the host bandwidth $12t=11\ev$ (such that $U=1.79\ev$ and $V=0.51\ev$) we obtain $T_K \approx 57\kel$, consistent with established results\cite{00:UjsaghyZawadowski,06:CastroNetoJones,04:WahlSchneider} for Co atoms on a Cu surface. As seen from Fig.~\ref{fig:Gimp} (center), the Kondo effect results in a large imaginary part $-t\text{Im}~G_{d}(\omega)\sim 6t^{2}/(\pi V^2)\simeq 6.2$ for low energies $|\omega|\ll T_K$.

For the 2d square lattice, the low-energy divergence in the host density of states results in an enhanced Kondo temperature.\cite{13:MitchellBullaFritz,13:MitchellFritzPLK} In any Kondo phase, the pinning condition from Eq.~\ref{eq:specpin} still holds,\cite{Del:pt,14:TuckerLogan} implying that,
\begin{eqnarray}
\label{eq:Gimp2d}
-\text{Im}~G_{d}(\omega) ~\overset{|\omega|\rightarrow 0}{\sim}~ \frac{2\pi t}{V^2} \big [ \ln(16 t/|\omega|) \big ]^{-1} \;,
\end{eqnarray}
which decays logarithmically at low energies. As confirmed in the left panel of Fig.~\ref{fig:Gimp}, the impurity spectrum $-\text{Im}~tG_{d}(\omega)$ therefore shows a maximum at $|\omega|\sim T_K$. With the same parameters as the 3d cubic system, we now obtain a much higher $T_K\approx 584\kel$.\cite{note:tkdef}

Finally, in the pseudogapped honeycomb lattice the Kondo effect is suppressed due to the depleted density of states near the Fermi level, and the local moment phase is stable for any $U/V^2$ at particle-hole symmetry.\cite{98:GonzalezBuxtonIngersent,13:FritzVojta} The impurity spectrum then takes the low-energy form
\begin{eqnarray}
\label{eq:Gimphc}
-\text{Im}~G_{d}(\omega) ~\overset{|\omega|\rightarrow 0}{\sim}~ \alpha' |\omega| 
\end{eqnarray}
(with $\alpha'$ a constant). This decay of $G_{d}(\omega)$ is demonstrated in the right panel of Fig.~\ref{fig:Gimp}. With $t=2.84\ev$, the honeycomb lattice models the $\pi/\pi^*$ bands of graphene;\cite{02:ReichMaultzsch} we use $U=0.704t$ and $V=0.493t$ as realistic impurity parameters obtained from ab-initio calculations for Co atoms on graphene.\cite{10:WehlingRosch} 

We now turn to the QPI for these three lattices.


\subsection{2d square lattice}
\label{sec:qpi_1imp_2d}
\begin{figure}[t]
\includegraphics[height=7cm]{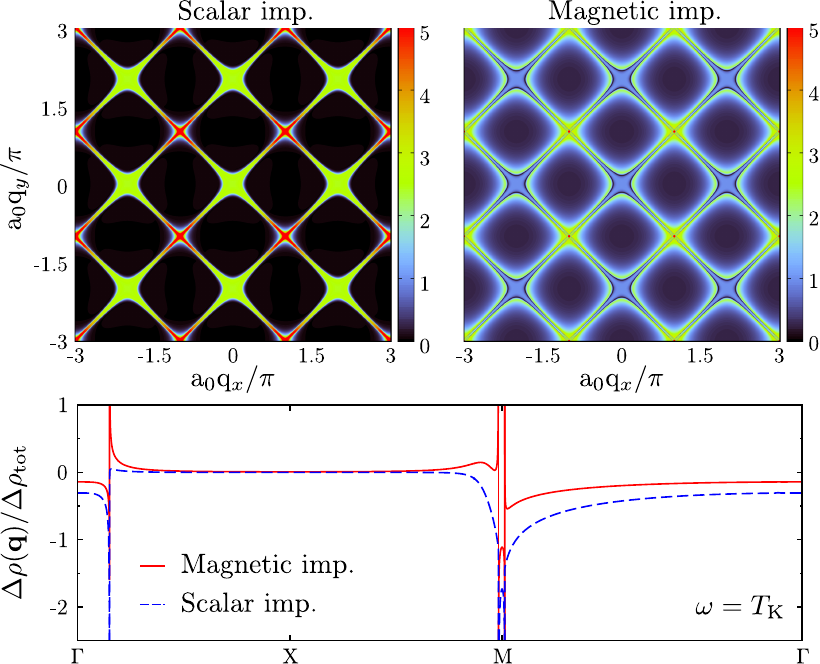}
\caption{QPI for a single impurity on the 2d square lattice at a scanning energy $\omega=0.055t \simeq T_{K}$. Upper panels compare the QPI maps $|\Delta\rho(\bold q)|$ for scalar (left) and magnetic (right) impurities; lower panel shows a Brillouin zone cut along the path $\Gamma\to\mathrm{X}\to\mathrm{M}\to\Gamma$, plotted as $\Delta\rho(\bold q)/\Delta\rho_{\mathrm{tot}}$. Symmetry points defined as $\bold q_{\Gamma} = \mathbf{0}$, $\bold q_{\mathrm{X}}=\mathbf{A}_1$, $\bold q_{\mathrm{M}}=\mathbf{A}_1+\mathbf{A}_2$ in terms of reciprocal lattice vectors $\mathbf{A}_1=2\pi/\mathrm{a}_0(1,0)$, $\mathbf{A}_2=2\pi/\mathrm{a}_0(0,1)$. Magnetic impurity parameters as in Fig.~\ref{fig:Gimp} such that $\omega=5\times 10^{-2}t=T_K$; and $v=0.5t$ for the scalar impurity.
}\label{fig:qpi2d}
\end{figure}

The QPI is obtained from Eq.~\ref{eq:qpi_1imp_cs_ps} or \ref{eq:qpi_1imp_cs_mag}, with $\Lambda(\bold q,\omega)$ for a single impurity given from Eq.~\ref{eq:lambda_def_sq} (with $\boldr_\alpha=\boldr_\beta=\bold 0$), 
\begin{equation}
\label{eq:lambda_1imp_2d} 
\Lambda(\bold q,\omega) = \int\limits_{1SBZ}\frac{d^2\mathrm{\bold{k}}}{\VBZ}~G^{0}(\boldk,\omega)G^{0}(\boldk-\bold q,\omega) \;.
\end{equation}
Fig.~\ref{fig:qpi2d} shows the absolute value of the QPI $|\Delta\rho(\bold q)|$ as a colour map in $\bold q$-space (upper panels), comparing the scalar impurity (left) with the magnetic impurity (right), at a fixed scanning energy $\omega \simeq T_K$, using the same parameters as Fig.~\ref{fig:Gimp}. The lower panel shows a cut across the Brillouin zone of $\Delta\rho(\bold q)/\Delta\rho_{\mathrm{tot}}$, where $\Delta\rho_{\mathrm{tot}} \equiv \Delta\rho_{\mathrm{tot}}(\omega)=\int_{1BZ}d^2q|\Delta\rho(\boldq,\omega)|$ is the total scattering amplitude at energy $\omega$.
For a single impurity, the topology in $\bold q$-space, on which we now focus, is completely determined by the host function 
$\Lambda(\bold q,\omega)$ (see Eqs.\ \ref{eq:qpi_1imp_cs_ps}, \ref{eq:qpi_1imp_cs_mag}).

\begin{figure}[t]
\includegraphics[height=7cm]{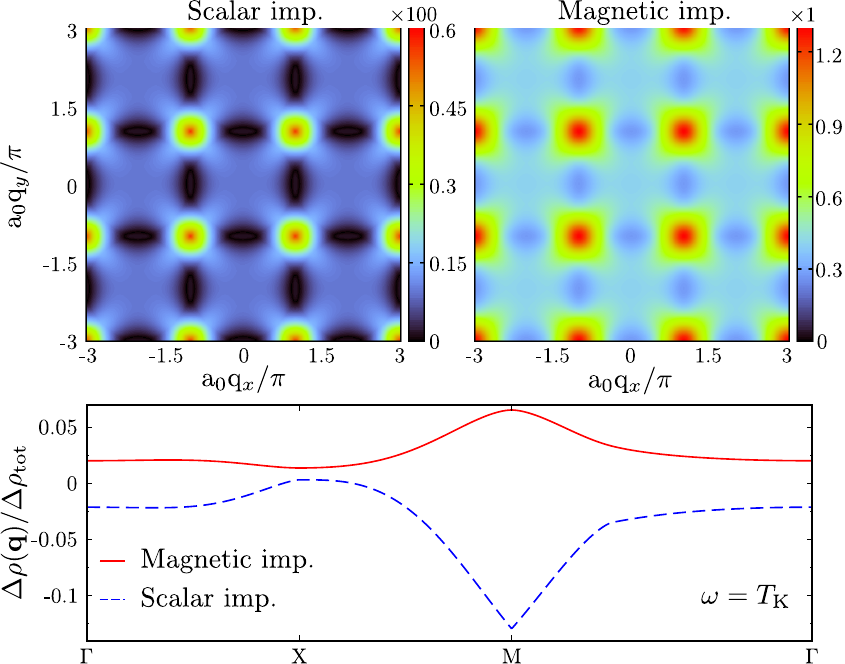}
\caption{QPI for an impurity on the (100) surface of a 3d cubic lattice. As Fig.~\ref{fig:qpi2d} but at $\omega=5\times10^{-3}t=T_K$.
}\label{fig:qpi3d}
\end{figure}

For scalar impurities in the Born limit, only the imaginary part, $\Lambda''(\bold q,\omega)$, plays a role (Eq.~\ref{eq:qpi_1imp_cs_ps}). For given scanning energy $\omega$, its structure gives rise to singular lines at $\bold q = \bold q^{*}(\omega)$ associated with the van Hove point of the 2d square lattice. Around the $\Gamma$ symmetry point, these lines are rectangular hyperbola,
\begin{equation}
\label{eq:qpi1imp_2d_gam}
\left (q^{*}_x\right )^2 - \left (q^{*}_y\right )^2 ~\overset{|\omega|\ll t}{\sim}~ \pm a_{\Gamma}(\omega)^2 ~~:~-\pi < q^{*}_{x,y} \le \pi \;, 
\end{equation}
with the dispersive properties controlled by $a_{\Gamma}(\omega)$; while around the M point 
\begin{equation}
\label{eq:qpi1imp_2d_m}
\left ( q^{*}_x \pm q^{*}_y\right )^2 ~\overset{|\omega|\ll t}{\sim}~  \left ( 2\pi -a_{\text{M}}(\omega) \right)^2 ~~:~-\pi < q^{*}_{x,y} \le \pi \;. 
\end{equation}
In $\bold q$-space, $\Lambda''(\bold q,\omega)$ is found to diverge as
\begin{equation}
\label{eq:qpi1imp_2d_div}
\Lambda''(\bold q,\omega) ~\overset{\bold q \rightarrow \bold q^{*}}{\sim}~ \big|\bold q - \bold q^{*}\big|^{-1/2} 
\end{equation}
when approaching a point $\bold q^{*}$ from the $\Gamma$ or M points. The region enclosed by these divergences is therefore characterized by high QPI scattering intensity --- see Fig.~\ref{fig:qpi2d} for the scalar impurity. $\Lambda''(\bold q,\omega)$ does not however diverge on approaching from X, and remains comparatively small in its vicinity. In fact, $\Lambda''(\bold q,\omega)$ is \emph{odd} in $\omega$ due to the exact symmetry $\Lambda(\bold q,\omega)=\Lambda(\bold q,-\omega)^{*}$.  The scalar impurity QPI $\Delta \rho (\bold q,\omega)\sim\omega$ thus vanishes at low energies away from the lines of divergence.

The situation is rather different for the magnetic impurity because both real and imaginary parts of $\Lambda(\bold q,\omega)$ are important (Eq.~\ref{eq:qpi_1imp_cs_mag}). Because $\Lambda'(\bold q,\omega)$ is even in $\omega$, residual QPI intensity around the X symmetry point persists even at low energies, due to finite $\Lambda'(\bold q,\omega=0)=b'_{\bold q}$. $\Lambda'(\bold q,\omega)$ also diverges logarithmically (as Eq.~\ref{eq:qpi1imp_2d_div}) on approaching the singular lines from X (it does not diverge in the vicinity of $\Gamma$ or M). As such, the QPI scattering intensity is enhanced around X for magnetic impurities. 

Further, as shown in the lower panel of Fig.~\ref{fig:qpi2d}, the \emph{sign} of $\Delta\rho(\bold q,\omega)$  can be reversed by the contribution from the second term in Eq.~\ref{eq:qpi_1imp_cs_mag}. This is a hallmark of scattering from magnetic impurities, where $\text{Im}~G_{d}(\omega)<0$ can become large due to the Kondo effect (see Fig.~\ref{fig:Gimp}). This leads to additional structure in the measurable $|\Delta\rho(\bold q)|$, not found in QPI for scalar impurities.


\subsection{3d cubic lattice with (100) surface}
\label{sec:qpi_1imp_3d}
The (100) surface of the 3d cubic lattice is again a square lattice; but `surface' quasiparticles are dephased by coupling to the bulk. This leads to the t-matrix Eq.~\ref{eq:tmeqn_kpar}, and for a single impurity
\begin{equation}
\label{eq:lambda_1imp_3d} 
\Lambda(\bold q,\omega) = \int\limits_{1SBZ}\frac{d^2\mathrm{\bold{k}}_{\parallel}}{\VBZ}~G_{\text{surf}}^{0}(\boldk_{\parallel},\omega)G_{\text{surf}}^{0}(\boldk_{\parallel}-\bold q,\omega) 
\end{equation}
with scattering vectors $\bold q \equiv \bold q_{\parallel}$ confined to the 2d surface (and the $G_{\text{surf}}^{0}(\boldk_{\parallel},\omega)$ given by Eq.~\ref{eq:G01d}).

The resulting QPI, shown in Fig.~\ref{fig:qpi3d}, does not contain the divergences arising in the 2d square lattice --- but remnants of this singular structure appear in broadened regions of enhanced scattering intensity around the M symmetry point in the cubic lattice. The global four-fold symmetry of the QPI evolves locally into a continuous rotational symmetry around this point, with $\Lambda(\bold q,\omega) \equiv \Lambda (|\bold q - \bold q_{\text{M}} |, \omega )$ for $|\omega|\ll t$. Further, the QPI for the scalar impurity (which only depends on $\Lambda''(\bold q,\omega)$) is distinctly conical, with 
\begin{equation}
\label{eq:imlambda3dsurf_m}
\Lambda''(\bold q,\omega)\overset{\bold q \rightarrow \bold q_{\text{M}}}{\sim} a''_{\text{M}}(\omega)+ b''_{\text{M}}(\omega) \left | \bold q - \bold q_{\text{M}}\right | + \mathcal{O} \left | \bold q - \bold q_{\text{M}} \right | ^2 \;,
\end{equation}
whereas the QPI for a magnetic impurity can become dominated by the quadratic term (see Fig.~\ref{fig:qpi3d}) since
\begin{equation}
\label{eq:relambda3dsurf_m}
\Lambda'(\bold q,\omega)\overset{\bold q \rightarrow \bold q_{\text{M}}}{\sim} a'_{\text{M}}(\omega)+ c'_{\text{M}}(\omega) \left | \bold q - \bold q_{\text{M}}\right |^2 + \mathcal{O} \left | \bold q - \bold q_{\text{M}} \right | ^4 \;.
\end{equation}

A striking feature of the QPI for the 3d cubic lattices is the difference in \emph{intensity} between scalar and magnetic impurities (note the rescaled color range in Fig.~\ref{fig:qpi3d}). 
There are two distinct reasons for this. First, $\Lambda''(\bold q,\omega)$ is odd in $\omega$,  whence $\Lambda''(\bold q,\omega)\sim \omega$ at low energies $|\omega|\ll t$. For the scalar impurity, Eq.~\ref{eq:qpi_1imp_cs_ps} implies that the QPI, $\Delta \rho (\bold q) \sim \omega$, is therefore also small. By contrast, the QPI for a magnetic impurity (Eq.~\ref{eq:qpi_1imp_cs_mag}) has a contribution from $\Lambda'(\bold q,\omega)$, which remains finite as $\omega\rightarrow 0$.

Second, in the case of magnetic impurities, the Kondo effect produces a spectral resonance in $\text{Im}~G_{d}(\omega)$ of width $T_K$ that does not decay at low energies (cf. Eq.~\ref{eq:Gimp2d} and Fig.~\ref{fig:Gimp}). In consequence, the QPI is considerably more intense at low energies for magnetic impurities than scalar impurities in standard flat-band metallic systems.


\subsection{2d honeycomb lattice}
\label{sec:qpi_1imp_hc}

The 2d honeycomb lattice, modelling the $\pi$ and $\pi^*$ bands in graphene within a nearest-neighbor tight-binding picture,\cite{02:ReichMaultzsch} generates richer structure in the QPI,\cite{10:WehlingRosch,11:WehlingKatsnelson,11:SimonBena} due to the bipartite nature of the lattice and the low-energy pseudogapped density of states (Eq.~\ref{eq:dos}). 

A single impurity coupled to a single honeycomb site (on sublattice $\gamma'=A$ or $B$) lowers the symmetry by breaking the centrosymmetry of the lattice. The single-impurity QPI is therefore obtained from Eq.~\ref{eq:QPI_Q} and Eq.~\ref{eq:Q_1imp}, with $\Lambda(\bold q,\omega)$ comprising contributions from both sublattices,
\begin{equation}
\label{eq:lambda_1imp_hc} 
\Lambda(\bold q,\omega)=\sum_{\gamma}\Lambda_{\alpha\alpha}^{\gamma,\gamma',\gamma'}(\bold q,\omega) \; .
\end{equation}
$\Lambda_{\alpha\alpha}^{\gamma,\gamma',\gamma'}(\bold q,\omega)$ itself is given by Eq.~\ref{eq:hc_lambda_def}, and depends on the phase $\phi$ defined in Eq.~\ref{eq:hc_phase}. This phase has a marked qualitative effect on the resulting QPI, and cannot be neglected. For centrosymmetric lattices, the QPI is periodic across the first Brillouin zone because $\Lambda(\bold q+n\textbf{A}_i,\omega) = \Lambda(\bold q,\omega)$ with integer $n$ for any reciprocal lattice vector $\textbf{A}_i$. But in the honeycomb lattice
\begin{equation}
\label{eq:hc_periodicity}
\Lambda(\bold q+3n\textbf{A}_i,\omega) = \Lambda(\bold q,\omega) \;,
\end{equation}
arising because $\phi(\bold k+n\textbf{A}_i)=\phi(\bold k)+\exp[2n\pi\I/3]$.~\cite{note:FTP} As such, the period of the QPI is enlarged to include the third Brillouin zone. 

\begin{figure}[t]
\includegraphics[height=7cm]{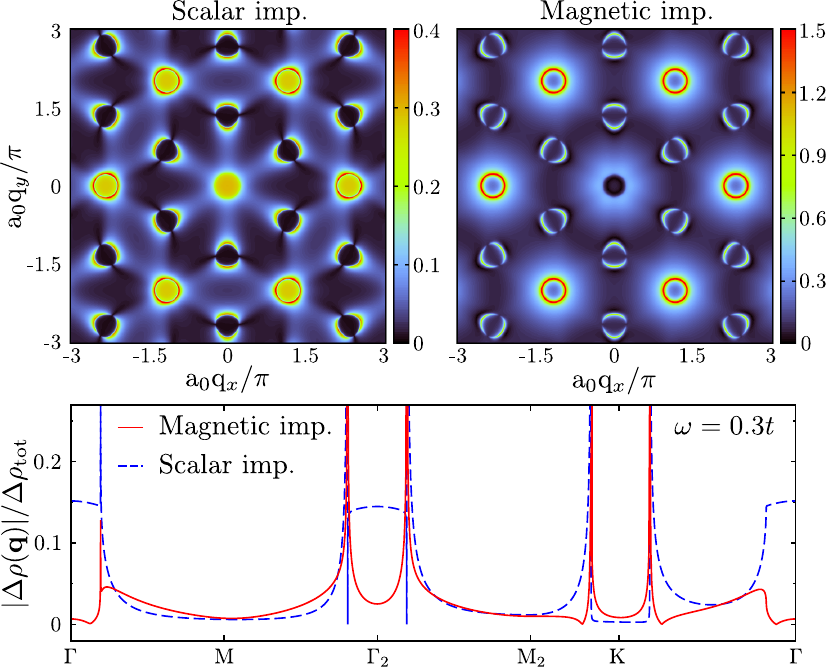}
\caption{QPI for a single impurity on the honeycomb lattice. As Fig.~\ref{fig:qpi2d} but at $\omega=0.3t$ and with impurity parameters from Fig.~\ref{fig:Gimp}. The Brillouin zone cut takes the path $\Gamma\to\mathrm{M}\to\Gamma_2\to\mathrm{M}_2\to\mathrm{K}\to\Gamma$, where $\bold q_{\Gamma}=\bold 0$, $\bold q_{\text{M}}=\tfrac{1}{2}\textbf{A}_1$, $\bold q_{\Gamma_2}=\textbf{A}_1$, $\bold q_{\text{M}_2}=\textbf{A}_1+\tfrac{1}{2}\textbf{A}_2$, $\bold q_{\text{K}}=\tfrac{2}{3}\textbf{A}_1+\tfrac{1}{3}\textbf{A}_2$, in terms of reciprocal lattice vectors $\textbf{A}_{1,2}=2\pi/\text{a}_0(\tfrac{1}{\sqrt{3}},\pm 1)$.
}\label{fig:qpihc}
\end{figure}

One consequence of this is the inequivalence of intravalley scattering at the $\Gamma$ and $\Gamma_2$ points (located at $\bold q_{\Gamma}=\bold 0$ and $\bold q_{\Gamma_2}=\textbf{A}_1$ respectively), where $\Delta\rho(\bold q_{\Gamma_2},\omega)=\Delta\rho(\bold q_{\Gamma},\omega)\times \exp[-\pi\I/3]$. At low energies $|\omega|\ll t$, these points are surrounded by singular lines in $\Lambda(\bold q,\omega)$ at $\bold q=\bold q^{*}$, where
\begin{equation}
\label{eq:hc_divgam}
|\bold q- \bold q^{*}| = d_{\Gamma}(\omega)  \;,
\end{equation}
giving rise to a circular derivative discontinuity in the QPI around $\bold q_{\Gamma}$ and $\bold q_{\Gamma_2}$. The dispersive properties are controlled by $d_{\Gamma}(\omega)=d_{\Gamma_2}(\omega)$, discussed further in the following subsection. For the scalar impurity at low energies $|\omega|\ll t$, both points are surrounded by flat regions of high scattering intensity,
\begin{equation}
\label{eq:hc_qpigam}
|\Delta\rho(\bold q,\omega)| ~\overset{|\bold q|<|\bold q^{*}|}{\sim}~ b_{\Gamma}  \;,
\end{equation}
with $b_{\Gamma}=b_{\Gamma_2}$ independent of scanning energy $\omega$. However, the local environment of the $\Gamma$ and $\Gamma_2$ points is different. The immediate vicinity of the $\Gamma_2$ point possesses a continuous rotational symmetry, with divergences in the QPI along the entire singular line $\bold q=\bold q^{*}$,
\begin{equation}
\label{eq:qpi1imp_hcdiv}
\Lambda(\bold q,\omega) ~\overset{\bold q \rightarrow \bold q^{*}}{\sim}~ \ln^2\big|\text{a}_0\bold q - \text{a}_0\bold q^{*}\big| \;.
\end{equation}
By contrast, a lower six-fold symmetry is found around the $\Gamma$ point as $\omega\rightarrow 0$ due to divergent \emph{points} arising only when $\bold q^{*}\times \boldsymbol \delta=\bold 0$, with $\boldsymbol \delta= \textbf{A}_1$, $\textbf{A}_2$ and $(\textbf{A}_1+\textbf{A}_2)$. 

Complex features in the QPI also appear in the vicinity of the K symmetry points due to intervalley scattering, and are again enclosed by singular lines, denoted $\bold q^{*}$. These features possess only reflection symmetry about the line $(\bold q_{\Gamma}-\bold q_{\text{K}})$, the continuous rotational symmetry being lifted by the underlying phase texture (itself arising because the impurity couples to a single sublattice). The line of divergence along $\bold q^{*}$ is intersected by a perpendicular nodal line at pinch-points where $(\bold q^{*}-\bold q_{\text{K}})\cdot(\bold q_{\Gamma}-\bold q_{\text{K}})=0$. For the scalar impurity, scattering is forbidden within the region around the K point enclosed by the singular lines. These features are seen clearly in the QPI map and cuts for the scalar impurity presented in Fig.~\ref{fig:qpihc}. At higher scanning energies, trigonal warping sets in, giving rise to a local three-fold point symmetry around $\bold q_{\text{K}}$.

For the magnetic impurity, the relative weight of $\Lambda'(\bold q,\omega)$ and $\Lambda''(\bold q,\omega)$ in the QPI depends on the complex t-matrix, $T(\omega)$, which evolves with scanning energy $\omega$ (see Fig.~\ref{fig:Gimp}). Importantly, this can lead to distinctive features in the measurable QPI, $|\Delta \rho(\bold q,\omega)|$. Accidental cancellation of terms in Eq.~\ref{eq:QPI_Q} can produce `dark spots' of suppressed scattering in the QPI. An example is shown Fig.~\ref{fig:qpihc}, where $|\Delta \rho(\bold q,\omega)|\simeq 0$ for $|\bold q|<|\bold q^{*}|$ in the vicinity of the $\Gamma$ point. In contrast to the scalar impurity case (Eq.~\ref{eq:hc_qpigam}), the QPI in general depends on $\omega$ and varies with $\bold q$ in the vicinity of the $\Gamma$ and $\Gamma_2$ points when magnetic impurities are present. Indeed, magnetic impurities also induce scattering near the K point.


\begin{figure*}[t]
\includegraphics[width=14cm]{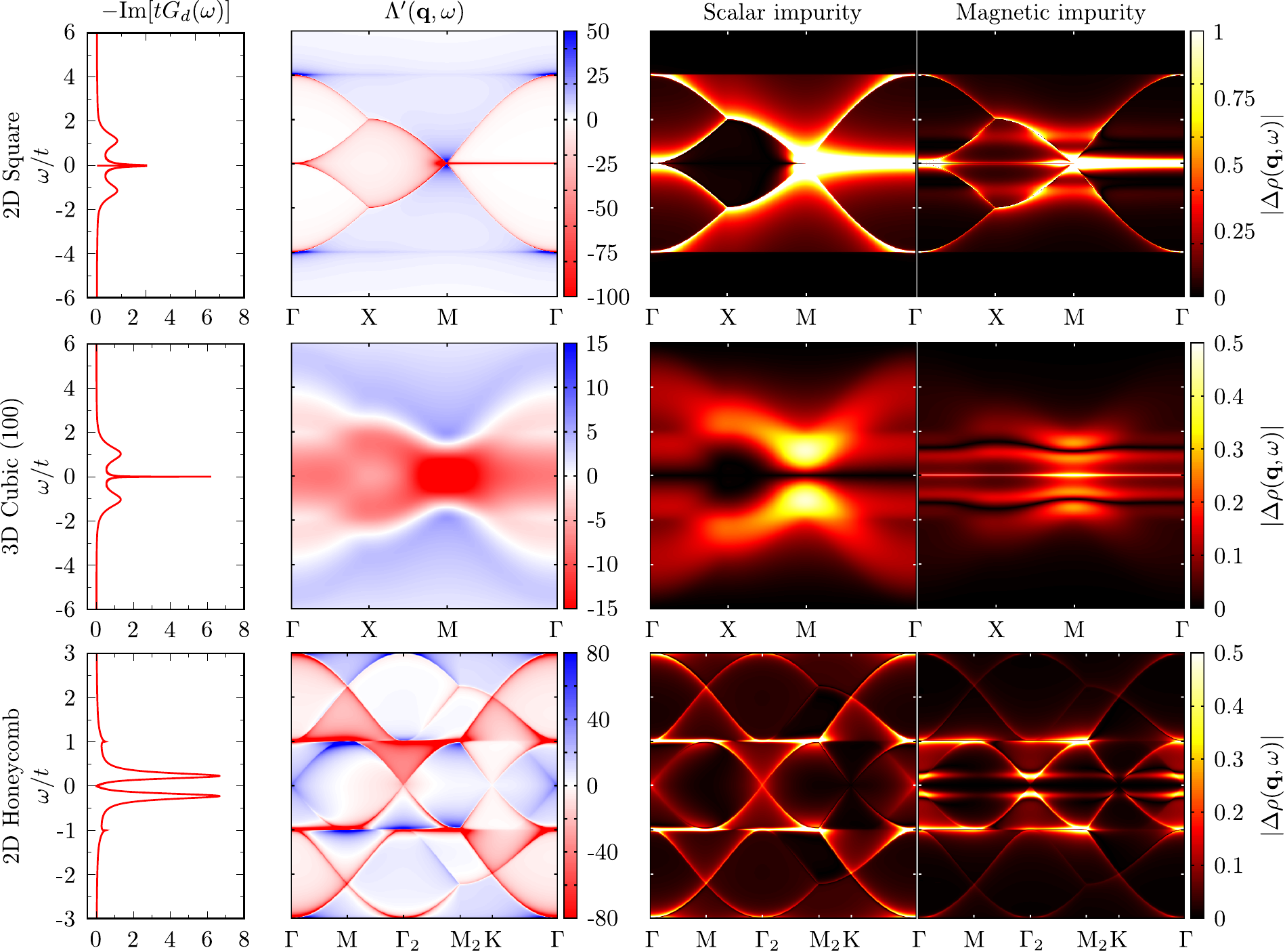}
\caption{QPI dynamics for a single impurity on various lattices: 2d square, 3d cubic with (100) surface, honeycomb (top, middle, and bottom row panels respectively). 
Right column panels: colour plots of the QPI $|\Delta\rho(\bold q,\omega)|$ across a Brillouin zone cut as a function of scanning energy $\omega/t$, comparing scalar and magnetic impurities. 
Center column panels show a colour plot of the host function $\Lambda'(\bold q,\omega)$ over the same $\bold q$-cut and energies. 
Left column panels show the spectral function for a magnetic impurity, $-\text{Im}[tG_{d}(\omega)]$ vs $\omega/t$, calculated via NRG at $T=0$. 
Magnetic impurity parameters as in Fig.~\ref{fig:Gimp}. Scalar impurity $v=0.15t$.
}\label{fig:qpidyn}
\end{figure*}

\section{Characteristic Kondo physics\\in the QPI}
\label{sec:characteristicKondo}

\subsection{Scanning-energy dependence}
\label{sec:dynamical}
We turn now to dynamical features of the QPI for the three lattices, comparing scalar and magnetic impurities. Numerically-exact results which exemplify the key physics are presented in Fig.~\ref{fig:qpidyn}

For scalar impurities, the scanning-energy dependence of the QPI is due entirely to the $\omega$-dependence of $\Lambda(\bold q,\omega)$, which characterizes the clean host lattice. The real part of this function is plotted as a colour map in the center column panels of Fig.~\ref{fig:qpidyn} (the real and imaginary parts are related by Hilbert transformation).

For magnetic impurities, QPI dynamics result from both $\Lambda(\bold q,\omega)$ and the impurity Green function $G_{d}(\omega)$, whose spectrum is plotted in the left column panels of Fig.~\ref{fig:qpidyn} (see also Fig.~\ref{fig:Gimp} and Eqs.~\ref{eq:specpin}--\ref{eq:Gimphc} for the detailed low-energy behavior). The nontrivial scanning-energy dependence of the QPI reflects the rich structure of the underlying quantum impurity problem.

For the 2d square lattice, divergences in $\Lambda(\bold q,\omega)$, described by Eqs.~\ref{eq:qpi1imp_2d_gam} and \ref{eq:qpi1imp_2d_m}, give rise to lines of intense scattering in the QPI. The dispersive properties of these features are controlled at low energies $|\omega|\ll t$ by $a_{\Gamma}(\omega)$ and $a_{\text{M}}(\omega)$, which are related by continuity at the edge of the Brillouin zone through $a_{\text{M}}(\omega) = a_{\Gamma}(\omega)^2/(2\pi)$. We find linear dispersion of the divergent features around the M symmetry point, implying
\begin{eqnarray}
\label{eq:qpi1imp_2d_ab}
a_{\text{M}}(\omega) \sim |\omega| \qquad ; \qquad
a_{\Gamma}(\omega) \sim |\omega|^{1/2} \;,
\end{eqnarray}
as confirmed directly in the upper panels of Fig.~\ref{fig:qpidyn}.

As $\omega\rightarrow 0$, the divergences are confined to the line $q^{*}_x=q^{*}_y$ connecting $\Gamma$ and M symmetry points, whence 
\begin{equation}
\label{eq:qpi1imp_2d_gam_m}
\Lambda (q_x=q_y,\omega)~\overset{|\omega|\ll t}{\sim}~a_{\bold q}\delta(\omega) + b_{\bold q}|\omega|+\I c_{\bold q}\frac{1}{\omega}+... \;.
\end{equation}
The QPI itself thus diverges along this line with the universal asymptotic form,
\begin{equation}
\label{eq:qpi1imp_2d_qpidiv}
|\Delta\rho(q_x=q_y,\omega)|~\overset{|\omega|\rightarrow 0}{\sim}~\begin{cases}
\left | \frac{1}{\omega} \right | \qquad &:~\text{scalar}  \\
\frac{1}{|\omega|\ln^2(16t/|\omega|)} \qquad &:~\text{mag.} \;,
\end{cases}
\end{equation}
where $\text{Re}~G_{d}(\omega\rightarrow 0) \sim [\ln(16t/|\omega|)]^{-2}$ is used in the case of the magnetic impurity (obtained by Hilbert transform of Eq.~\ref{eq:Gimp2d}). The divergence is thus sharper along the M-$\Gamma$ line for magnetic impurities, as evident in Fig.~\ref{fig:qpidyn}.

Away from this divergent line (e.g. along the cut $\Gamma\rightarrow X \rightarrow M$ in Fig.~\ref{fig:qpidyn}), the QPI is characterized by vanishing scattering intensity at low scanning energies due to
\begin{equation}
\label{eq:qpi1imp_2d_x}
\Lambda (q_x\ne q_y,\omega)~\overset{|\omega|\rightarrow 0}{\sim}~\tilde{a}_{\bold q}+\tilde{b}_{\bold q}|\omega|+\I \tilde{c}_{\bold q}\omega\ln|\omega/t| + ... \;,
\end{equation}
giving rise to the asymptotic behavior of the QPI,
\begin{equation}
\label{eq:qpi1imp_2d_qpinondiv}
|\Delta\rho(q_x\ne q_y,\omega)|~\overset{|\omega|\rightarrow 0}{\sim}~\begin{cases}
\omega \ln|\omega/t| \qquad &:~\text{scalar}  \\
\frac{1}{\ln(16t/|\omega|)} \qquad &:~\text{mag.} 
\end{cases}
\end{equation}
As a result, the QPI intensity for the magnetic impurity decays much more slowly than that of the scalar impurity as the scanning energy is reduced in the vicinity of the X symmetry point --- see Fig.~\ref{fig:qpidyn}.

At higher scanning energies $|\omega|\sim U/2$, the `Hubbard satellites' in the spectral function of the magnetic impurity give rise to enhanced scattering near the M and X points. These features are not of course present in the scalar impurity QPI, and are as such one signature of strong electron correlations in magnetic impurities. 

For the 3d cubic lattice the dynamics are rather different, for two reasons: the host function $\Lambda (\bold q,\omega)$ does not contain divergences, and the magnetic impurity Green function does not vanish at low energies because the host density of states is essentially flat for $|\omega|\ll t$.

For a scalar impurity on the cubic (100) surface, the QPI intensity vanishes everywhere at low energies $|\omega|\rightarrow 0$ because $\Lambda'' (\bold q,\omega)=-\Lambda'' (\bold q,-\omega)$ is odd in $\omega$ --- see center row panels of Fig.~\ref{fig:qpidyn}. By contrast, the Kondo effect gives rise to enhanced scattering at temperatures and energies $\ll T_K$; this gives rise to a large finite QPI intensity for $\omega\ll T_K$. This is the typical behavior expected for magnetic impurities in standard metallic systems. 

Finally, we consider QPI dynamics on the the honeycomb lattice. In the case of the scalar impurity, the region of intense intravalley scattering around $\Gamma$ and $\Gamma_2$ described by Eq.~\ref{eq:hc_qpigam} disperses linearly at low energies according to Eq.~\ref{eq:hc_divgam}, with $d_{\Gamma}(\omega)\sim \omega$. As $\omega\rightarrow 0$, the only divergent points are at $\bold q^*=\bold q_{\Gamma}$ and $\bold q_{\Gamma_2}$. At $\bold q=\bold q_{\text{K}}$, the QPI $\Delta\rho(\bold q_{\text{K}},\omega)=0$ is identically zero for any $|\omega|<t$. These features are shown in the lower panels of Fig.~\ref{fig:qpidyn}. 

The QPI for the magnetic impurity near these points shows intense scattering at energies $|\omega|\sim U/2$, corresponding to the Hubbard satellites in the impurity spectral function due to charge fluctuations. At lower energies, however, $G_{d}(\omega)$ vanishes linearly according to Eq.~\ref{eq:Gimphc} due to the host LDOS which also vanishes linearly at low energies. Importantly, the Kondo effect is suppressed at particle-hole symmetry, and the local moment phase is always stable for any interaction strength.\cite{98:GonzalezBuxtonIngersent,13:FritzVojta} Electron correlations give rise to the nontrivial spin-flip scattering typical of such degenerate non-Fermi liquid phases.\cite{98:GonzalezBuxtonIngersent,14:TuckerLogan} As a result the QPI for the magnetic impurity vanishes everywhere at low energies, according to
\begin{eqnarray}
\label{eq:qpi1imp_2d_honey}
|\Delta\rho(\bold q,\omega )|~\overset{|\omega|\rightarrow 0}{\sim}~
\begin{cases}
|\omega|\ln^3\big|\omega/t \big| \qquad &:~\bold q=\bold q_{\Gamma_{(2)}} \\
0 \qquad &:~\bold q=\bold q_{\text{K}}  \\
|\omega|\ln\big|\omega/t \big|  \qquad &:~\text{elsewhere} \;.
\end{cases}
\end{eqnarray}


\subsection{Universality}
\label{sec:universality}
We focus now on a single magnetic impurity on the (100) surface of a 3d cubic lattice --- the case most relevant to standard metallic systems where the host density of states becomes essentially flat at low energies (see Eq.~\ref{eq:dos}). The Kondo effect is operative in such systems,\cite{hewson} with a spectral resonance setting in on temperature/energy scales $\sim T_K$ --- see center panel of Fig.~\ref{fig:Gimp}. This resonance embodies enhanced spin-flip scattering, which screens the impurity local moment dynamically. Importantly, all physical properties depend only on the single emergent scale $T_K$ at low temperatures/energies, reflecting the \emph{universal} RG flow between local moment and strong coupling fixed points.\cite{hewson}

At sufficiently low temperatures $T\ll T_K$, the impurity Green function $G_{d}(\omega)$ is a universal function of $\omega/T_K$ on all energy scales $|\omega|\ll \min(t,V^2/U)$ --- not only for $|\omega|\ll T_K$ where strong coupling Fermi liquid behavior Eq.~\ref{eq:specpin} holds, but also for $|\omega|\gg T_K$ where local moment physics dominates the dynamics. In that case, the impurity spectral function takes the asymptotic form,\cite{02:GlossopLogan}
\begin{eqnarray}
\label{eq:GimpLM}
-\text{Im}~ G_{d}(\omega) ~\overset{|\omega|\gg T_K}{\sim}~ \frac{1}{1+a\ln^2|b\omega/T_K|} \;,
\end{eqnarray}
with $a,b=\mathcal{O}(1)$ constants. This behavior for $T_K\ll |\omega| \ll \min(t,V^2/U)$ is universal because the hybridization function $\text{Im}~\Gamma(\omega)$ is essentially constant for $|\omega| \ll \min(t,V^2/U)$ on the cubic lattice (and $\text{Re}~\Gamma(\omega)\sim \omega$).

Similarly, the real part of the host function $\Lambda'(\bold q, \omega)$ becomes constant on energy scales $|\omega|\ll t$, while $\Lambda'(\bold q, \omega)\sim \omega$. In the scaling limit $T_K\rightarrow 0$, $\text{Im}~ G_{d}(\omega)$ thus controls the energy-dependence of the QPI in the universal regime (see Eq.~\ref{eq:qpi_1imp_cs_mag}). In practice, non-universal contributions are negligible for finite $T_K\ll \min(t,V^2/U)$.

\begin{figure}
\includegraphics[width=6.5cm]{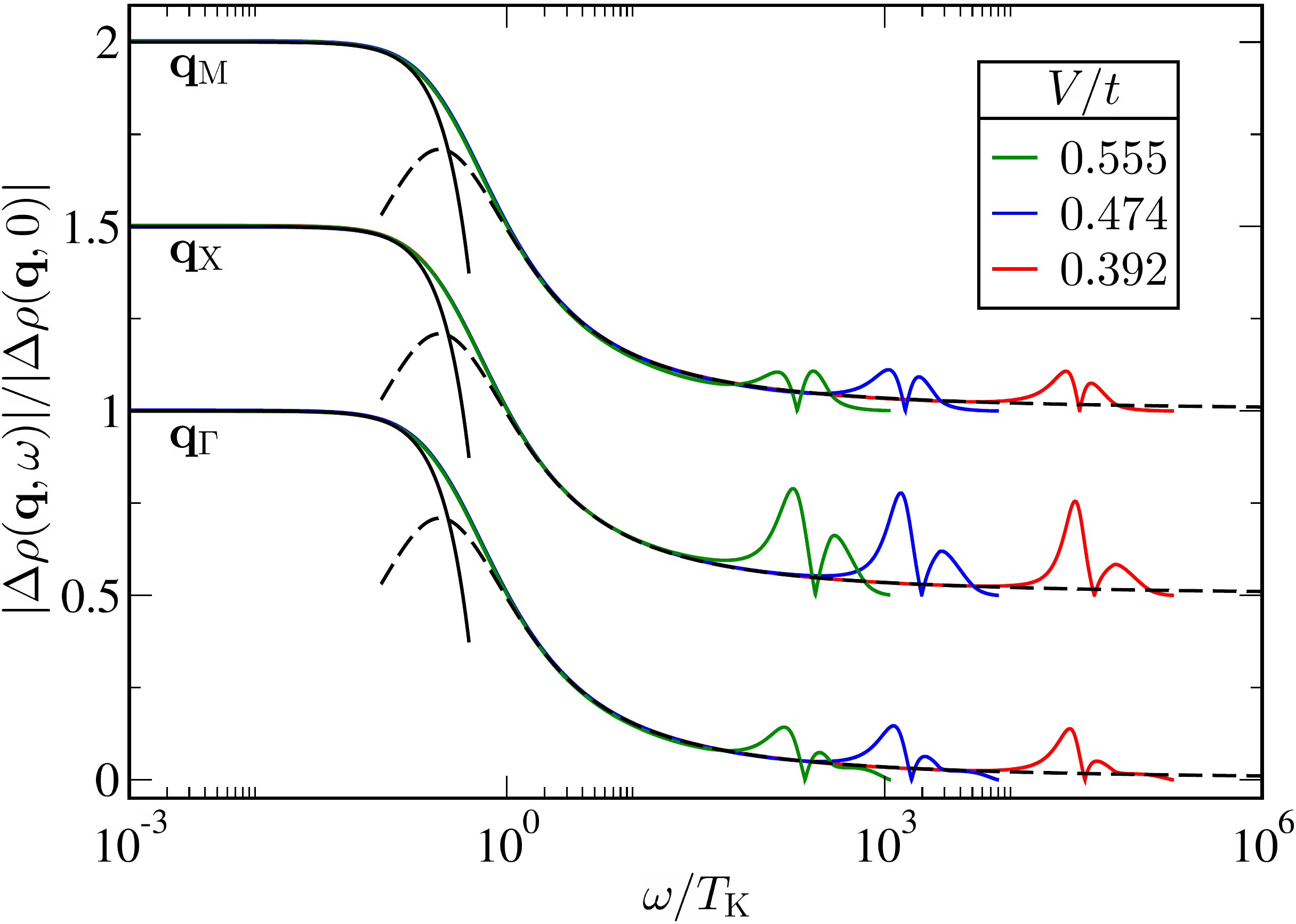}
\caption{QPI $|\Delta\rho(\boldq,\omega)|/|\Delta\rho(\boldq,0)|$ vs $\omega/\tk$ for a magnetic impurity on the 3d cubic (100) surface for scattering vectors $\bold q=\bold q_{\Gamma}$, $\bold q_{\text{X}}$ and $\bold q_{\text{M}}$ (vertically offset by $0.5$ for clarity), plotted for a range of impurity parameters: $V/t=0.555$, $0.474$, $0.392$ with fixed $U=1.95t$, corresponding to $\tk/t=5\times10^{-3}$, $7\times10^{-4}$ and $3\times10^{-5}$. Eq.~\ref{eq:specpin} (solid line) and Eq.~\ref{eq:GimpLM} (dashed) describe $|\omega|\ll T_K$ and $\gg T_K$ asymptotes.
}\label{fig:univ} 
\end{figure}

In consequence, the entire QPI develops a universal scanning-energy dependence at low energies and temperatures. This means that magnetic impurities with different interaction strengths $U$ and couplings $V$ on different metallic substrates give the same normalized low-temperature/energy QPI $|\Delta\rho(\bold q,\omega)|/|\Delta\rho(\bold q,0)|$ when plotted vs rescaled scanning-energy $\omega/T_K$, for any scattering vector $\bold q$. Rescaled experimental QPI data for different systems should thus collapse onto a part of this universal curve, providing the unambiguous signature of scattering from magnetic impurities. This scaling collapse is demonstrated for the 3d cubic lattice in Fig.~\ref{fig:univ}. Departure from universality is governed by the onset of $\omega$-dependence in $\Lambda(\boldq,\omega)$, which is distinct for each $\boldq$; for the experimentally relevant parameters used, this is found to occur at 
$|\omega| \gtrsim20\tk \simeq 0.1t$.

\begin{figure}
\includegraphics[width=8cm]{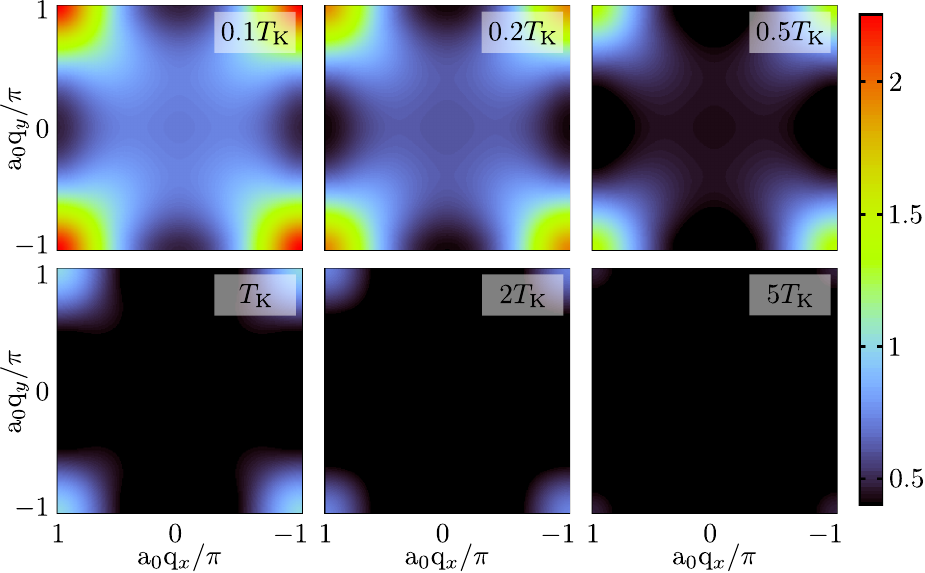}
\caption{\label{fig:thermalMaps}Simulated experimental FT-STS measurements $|\Delta\rho_\text{meas}(\mathbf{q},\omega=0.1\tk)|$ for a single magnetic impurity (parameters as in \fref{fig:Gimp}, such that $\tk=5\times10^{-3}t$) at fixed scanning energy, plotted across the 1BZ for a series of temperatures, $T/\tk= 0.1$, 0.2, 0.5, 1, 2 and 5.}
\end{figure}

\subsection{Thermal effects}
\label{sec:thermal}
So far we have considered $T=0$, appropriate in practice when $T\lesssim\tk\ll t$, such that temperature is the smallest energy scale in the problem. This regime is relevant, as typical STM experiments are conducted at $\sim5\kel$;\cite{04:WahlSchneider} but studies at higher $T$ may also be performed\cite{06:highTSTM} to investigate the change in QPI upon increasing temperature through $\tk$, and beyond.

Non-interacting conduction electrons and uncorrelated impurities (e.g.\ the scalar impurity) in practice have $T$-independent electronic structure, and thus QPI. By contrast, electronic correlations of the magnetic adatom exhibit a strong $T$-dependence, entering the t-matrix via the impurity Green function. As $T$ is increased in metallic systems, the Kondo singlet is broken, destroying the Kondo resonance on the scale $T\gtrsim\tk$ and resulting in local-moment physics.\cite{hewson} This results in a quite dramatic change in the t-matrix, and hence QPI, on increasing $T$ through $\tk$.

In addition to this interaction-driven $T$-dependence, the local tunneling current measured in STS is weakly $T$-dependent due to thermal excitation of conduction electrons; at finite-$T$, the differential conductance 
\begin{align}\label{eq:finiteTcond}
\frac{dI}{dV}(\mathbf{r}_{i},\omega=eV,T) \propto \int_{-\infty}^{\infty}d\varepsilon\;
\rho(\mathbf{r}_{i},\varepsilon,T) f'(\omega-\varepsilon,T),
\end{align}
where $f'(\omega-\varepsilon,T)=\tfrac{d}{d\omega} f(\omega-\varepsilon,T)$ and $f(x,T)=[1+\exp(x/T)]^{-1}$ is the Fermi function. \eref{eq:finiteTcond} represents the convolution of the LDOS ($T$-dependent only for the magnetic impurity) with a broadening kernel, controlled by $T$. 
The QPI measured via FT-STS, $\Delta\rho_\text{meas}(\boldq,\omega,T)$, is then related to the `pure' QPI by:
\begin{align}\label{eq:FTSTSvsQPI}
\Delta\rho_\text{meas}(\boldq,\omega,T) = \int_{-\infty}^{\infty}d\varepsilon\;\Delta\rho(\boldq,\varepsilon,T) f'(\omega-\varepsilon,T)
\end{align}

\begin{figure}
\includegraphics[width=8cm]{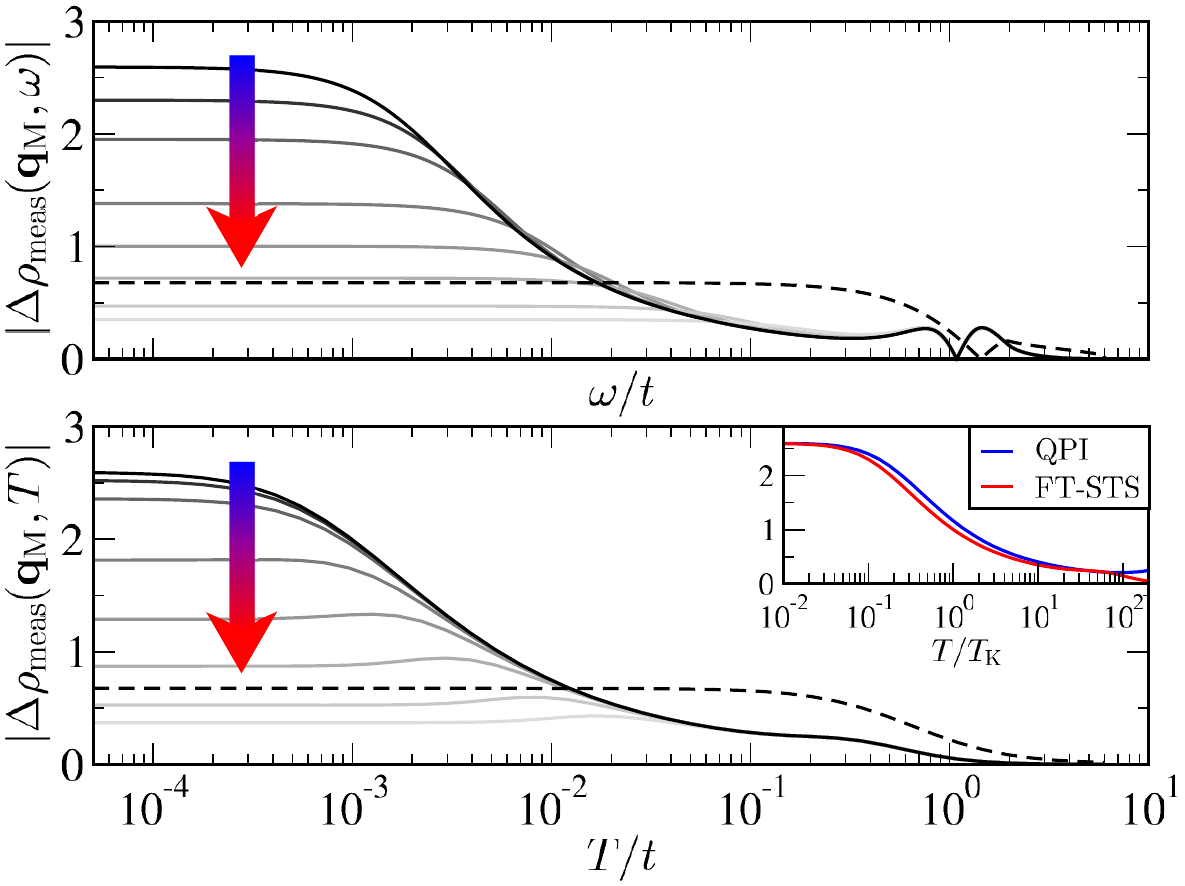}
\caption{\label{fig:thermalPts}QPI $|\Delta\rho_{\text{meas}}(\mathbf{q}_\text{M})|$ for a 
single magnetic impurity (parameters given in \fref{fig:Gimp}, such that $\tk=5\times10^{-3}t$) at the point 
$\mathbf{q}_{\mathrm{M}}=(\pi,\pi)$ in the 1BZ, shown as a function of $\omega$ ($T$) at a series of different $T$ ($\omega$) in the upper (lower) panel; such that $T/\tk~(\omega/\tk) = 0.01, 0.1, 0.2, 0.5, 1, 2, 5, 10$. Increasing $T$ from low to high is denoted by the arrow. The QPI due to a scalar impurity is also plotted (dashed line) for comparison. 
\emph{Lower panel inset:} the measured FT-STS (red) and the `pure' QPI (blue), \emph{vs} $T/\tk$, showing the same universal behavior in each case (up to a scale factor due to thermal broadening).}
\end{figure}

\fref{fig:thermalMaps} shows the thermal evolution of the QPI for an impurity embedded on the 3d cubic (100) surface. For the magnetic impurity, the magnitude of the QPI decreases substantially as $T$ increases through $\tk$, and the Kondo resonance is suppressed. By contrast the QPI for a scalar impurity has a much simpler $T$-dependence (entering only via the thermal broadening, Eq.~\ref{eq:FTSTSvsQPI}), with essentially no $T$-dependence for $T\ll t$, as depicted by the dashed line in \fref{fig:thermalPts}. The strong $T$-dependence of QPI is a characteristic signature of Kondo physics in systems with magnetic impurities.

The asymptotic low-$T$ expansion of the impurity Green function at particle-hole symmetry and $\omega =0$ (applicable in the regime $\omega\ll T\ll\tk$) is a universal function of the Kondo temperature,\cite{hewson}
\begin{align}\label{eq:lowTasymp}
\text{Im}\Gamma(\omega =0)\times \text{Im}\,G_d(\omega=0,T) \overset{T\ll\tk}{\sim} 1- \alpha_T(T/\tk)^2 +\dots
\end{align}
This leads to a close correspondence between the $\omega$- and $T$-dependence of the QPI (via the impurity Green function, comparing Eqs.\ \ref{eq:specpin},\ref{eq:lowTasymp}), as seen by comparison of the upper and lower panels in \fref{fig:thermalPts}. 

As such we expect to observe universal scaling in the $T$-dependence of the QPI for the magnetic impurity, analogous to that of the $\omega$-dependence. The comparison of the pure QPI and the thermally broadened FT-STS signal in the inset of \fref{fig:thermalPts} (inset) demonstrates this universal behavior, which is unaffected by the thermal broadening of the STM-measured conductance \eref{eq:FTSTSvsQPI} (up to a trivial scale factor).


\section{Interpretation of FT-STS}
\label{sec:interp}

\subsection{Finite-size effects}
\label{sec:finitesize}

In experiment, the surface LDOS $\Delta \rho(\bold r_i,\omega)$ is measured over an $L\times L$ plaquette using STM,\cite{02:HoffmanMcElroy} with the QPI obtained from Eq.~\ref{eq:QPI_def}. The $\mathbf{q}$-space resolution of the resulting QPI naturally depends on the real-space sample size. The `true' QPI is recovered as $L\rightarrow \infty$, obtained theoretically by the t-matrix approach.

We now consider explicitly the effects of finite sample size, by simulating the experimental protocol. The LDOS for these surface sites is calculated exactly using Eq.~\ref{eq:delta_rho}, with the non-local free Green functions obtained using the convolution method described in the appendix.

Fig.~\ref{fig:fs} shows a Brillouin zone cut through the QPI for a magnetic impurity on the cubic lattice (100) surface, computed in the $L\rightarrow \infty$ limit using the t-matrix approach (solid line). This true QPI is compared with results for the same system restricting to an $L\times L$ surface sample, with $L=100$ (crosses) and $L=10$ (diamonds). The true QPI is very well-approximated when $L=100$ is used (corresponding to a plaquette of side length $\sim 10^{2}$ \AA, as typical in experiment\cite{02:HoffmanMcElroy}). The $\bold q$-space resolution is also sufficient to capture accurately all features. Indeed, even for an extremely small sample region $L=10$, the accuracy is surprisingly good; although the discretization is severe. 

Reassuringly, the experimental protocol reproduces accurately the true QPI. A large sample size is however still needed to resolve sharp $\bold q$-space features; in 2d systems, the characteristic sharp cusps in the QPI in Figs~\ref{fig:qpi2d} and \ref{fig:qpihc} would require very large LDOS samples in real space.

\begin{figure}
\includegraphics[width=6.5cm]{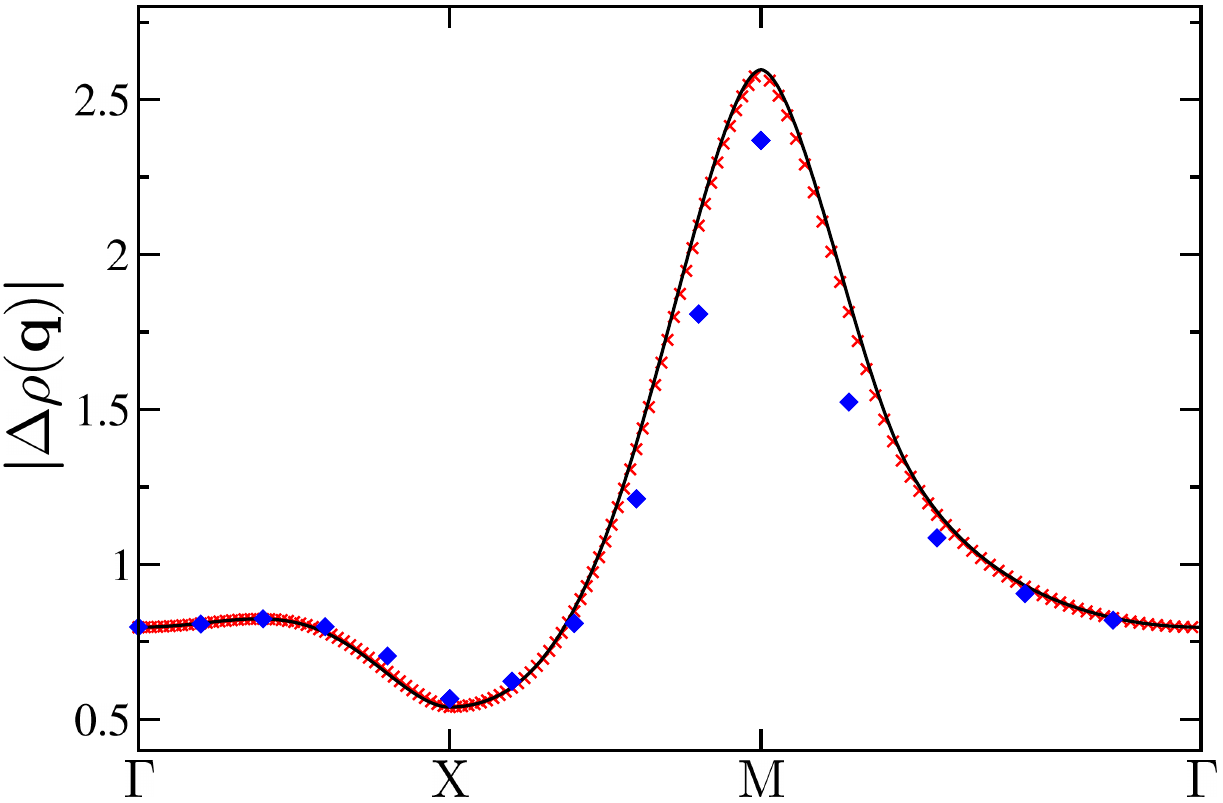}
\caption{\label{fig:fs}Brillouin zone cut of $|\Delta\rho(\boldq,\omega)|$ for a magnetic impurity on the 3d cubic (100) surface at $\omega=10^{-4}t$. Exact QPI (line) calculated via the t-matrix approach (Eq.~\ref{eq:qpi_1imp_cs_mag}), compared with the direct real-space approach (Eq.~\ref{eq:QPI_def}), sampling an $L\times L$ surface plaquette with $L=100$ (cross points) and $L=10$ (diamond points).
Impurity parameters as in Fig.~\ref{fig:Gimp}.}
\end{figure} 


\subsection{JDOS interpretation}
\label{sec:jdos}
The QPI $\Delta\rho(\bold q,\omega)$ is often interpreted (e.g.\ Refs.\ \onlinecite{02:HoffmanMcElroy},\onlinecite{11:SimonBena}) in terms of the joint density of states (JDOS, $J^{I}(\bold q,\omega)$), \emph{viz} $\Delta\rho(\bold q,\omega)\simeq J^{I}(\bold q,\omega)$ with
\begin{equation}
\label{eq:jdos}
J^{I}(\bold q,\omega)= \int\limits_{1SBZ} \frac{d^2\bold k}{\VBZ}~ \rho^0(\bold k, \omega)\rho^0(\bold k - \bold q,\omega) \
\end{equation}
and $\rho^0(\bold k,\omega)=-\tfrac{1}{\pi} \text{Im}~G^0(\bold k, \omega)$ the $\bold k$-resolved surface density of states at energy $\omega$. Contributions to the JDOS arise when quasiparticles on a constant energy contour with momenta $\bold k$ and $\bold k'$ are separated by $\bold q=\bold k-\bold k'$.

On heuristic grounds, it is usually argued that the amplitude of impurity-induced scattering from $\bold k$ to $\bold k'$ at energy $\omega$ can only be significant if there is a high density of quasiparticle states at both $\bold k$ \emph{and} $\bold k'$ (i.e. they have a large JDOS). The QPI at scattering vector $\bold q$ is the sum of all scattering processes where $\bold k'=\bold k-\bold q$. By assuming that the QPI is large when the JDOS is large, QPI patterns can be used to infer the JDOS and hence electronic properties of the clean host material. Experimental QPI data are typically interpreted in this way.\cite{02:HoffmanMcElroy,11:SimonBena} 

The usefulness and relative simplicity of the JDOS picture has motivated efforts  to connect rigorously the QPI and JDOS. Notably, the perturbative approach employed in Ref.~\onlinecite{07:SimonVonauAubel} makes the link by assuming a constant scattering amplitude and phase along the constant energy contour. But a faithful description of interfering scattering processes typically requires relative phase information; and the JDOS simply lacks information about overlap matrix elements between states in the impurity-coupled system. In consequence, the QPI may be small even when the JDOS is large (as may be verified explicitly). The JDOS picture then fails to capture the basic physics of the scattering --- as is known e.g.\ in graphene.\cite{11:SimonBena} We emphasize that $J^{I}(\bold q,\omega)$ cannot be derived from $\Delta\rho(\bold q,\omega)$ at a given $\bold q$ 
and $\omega$, in any controlled limit.

However for a simple scalar impurity on a centrosymmetric lattice, the QPI and JDOS are in fact related by Hilbert transformation. From Eqs.~\ref{eq:qpi_1imp_cs_ps} and \ref{eq:lambda_1imp_2d} (or e.g. Eq.~\ref{eq:lambda_1imp_3d}), the QPI in this case can be written as
\begin{align}
\label{eq:qpi_reGrho}
\Delta\rho(\bold q,\omega)= 2v\int\limits_{1SBZ} \frac{d^2\bold k}{\VBZ}~ \left[\text{Re}~G^0(\bold k, \omega)\right]\rho^0(\bold k - \bold q,\omega) \;,
\end{align}
where we have exploited periodicity across the 1SBZ. We introduce the complex quantity $J(\bold q,\omega)$, defined as
\begin{subequations}
\label{eq:qpi_jdos}
\begin{align}
J(\bold q,\omega)&= -\frac{1}{\pi}\int\limits_{1SBZ} \frac{d^2\bold k}{\VBZ}~ G^0(\bold k, \omega)\rho^0(\bold k - \bold q,\omega)  \;,\\
&\equiv-\frac{1}{2v\pi }\Delta\rho(\bold q,\omega) + \I J^{I}(\bold q,\omega) \;,
\end{align}
\end{subequations}
such that the QPI and JDOS are Hilbert conjugates, being respectively the real and imaginary parts of $J(\bold q,\omega)$.

\begin{figure}
\includegraphics[width=6.5cm]{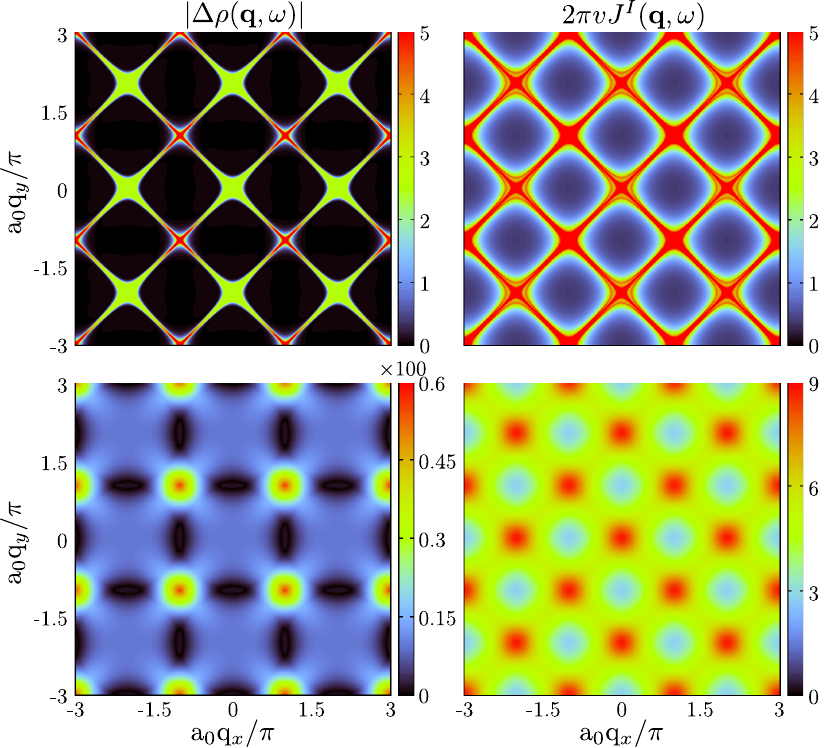}
\caption{\label{fig:qpiVjdos}Comparison of QPI, $|\Delta\rho(\boldq,\omega)|$ (left column), with the JDOS (right column) plotted as $2\pi v J^I(\boldq,\omega)$ (see Eq.~\ref{eq:qpi_jdos}); for the 2d square lattice (upper row) and (100) surface of 3d cubic lattice (lower row), for scanning energy $\omega$ and Born impurity scattering potential $v$ as in Fig.~\ref{fig:qpi2d} and \ref{fig:qpi3d} respectively. }
\end{figure}

The JDOS interpretation of the QPI may thus be roughly correct for dilute scalar impurities on centrosymmetric lattices, because the underlying $\bold q$-space topology of singular and nodal lines is the same for $J^{I}(\bold q,\omega)$ as it is for $\Delta\rho(\bold q,\omega)$, both being related to a single complex analytic function $J(\bold q,\omega)$. Nevertheless, even in this case the QPI is \emph{not} accessible directly from the JDOS at a given $\bold q$ and $\omega$: the \emph{entire} $\omega$-dependence of $J^{I}(\bold q,\omega)$ must be known to obtain $\Delta\rho(\bold q,\omega)$ by Hilbert transformation. By way of illustration, Fig.~\ref{fig:qpiVjdos} compares the JDOS to the QPI for the 2d square and (100) surface of the 3d cubic lattice; in the former case the two quantities are roughly similar, while in the latter the JDOS has significantly different $\boldq$-space structure and intensity.

We add that the above connection (Eq.\ \ref{eq:qpi_jdos}b) does \emph{not} hold in the case of non-centrosymmetric lattices or bipartite lattices, where Eq.~\ref{eq:qpi_reGrho} is inapplicable. The QPI, $\Delta\rho(\bold q,\omega)$, then depends on both real and imaginary parts of $J(\bold q,\omega)$ due to additional $\bold q$-dependent phase factors. This explains the failure of the JDOS interpretation in the case of impurities in graphene\cite{11:SimonBena} (see Sec.~\ref{sec:qpi_1imp_hc}). 

Finally, we emphasize that the JDOS and QPI are never related simply by Hilbert transformation when magnetic impurities are present. This is because the t-matrix is a complex dynamical object: the real and imaginary parts of $J(\bold q,\omega)$ are again mixed. Indeed the Hilbert transform, involving integration over all energy scales, necessitates a full knowledge of the impurity dynamics --- information simply not contained in the JDOS.


\section{Conclusion} 
\label{sec:conc}

We have studied theoretically the use of quasiparticle interference (QPI) --- measured in FT-STS experiments --- as a probe of magnetic adatoms on surfaces. Following a general formulation of the QPI due to an arbitrary distribution of impurities, we turned explicitly to single-impurities adsorbed on a range of host surfaces: the (100) surface of a 3d simple cubic lattice, and the 2d honeycomb and square lattices, in which the Fermi-level densities of states respectively embody standard metallic behavior, pseudogap behavior, and a divergence due to a Van Hove singularity. In all cases, the single-impurity QPI factorizes into a local scattering t-matrix, and a host response function $\Lambda(\boldq,\omega)$ at scanning energy $\omega =\ev$.

The scattering t-matrix for a magnetic impurity is simply related to the impurity Green function --- itself dependent on the host lattice --- and thus the rich dynamics due to electronic correlations is manifest in the QPI. Despite the local, momentum-independent nature of these correlations, the $\boldq$-space structure of the QPI is found to be qualitatively different from that of a simple scalar impurity due to non-trivial phase shifts associated with scattering from magnetic impurities, which reflect e.g.\ the Kondo effect.

The response function $\Lambda(\boldq,\omega)$ is also non-trivial, despite being a property of the free, non-interacting host. 
It displays significant structure in $\boldq$-space, symptomatic of the symmetry and dimensionality of the host,
but its energy-dependence becomes featureless for $|\omega|\ll t$ (with $t$ the intersite lattice hopping).
By contrast, the Kondo physics due to a magnetic impurity is controlled by an emergent scale $\tk\ll t$, and so scattering becomes strongly energy-dependent at low energies. Indeed, the QPI exhibits \emph{universal} scaling in terms of $\omega/\tk$ and $T/\tk$ --- a characteristic hallmark for systems containing magnetic impurities. Conversely, the QPI for systems containing scalar impurities has no energy or temperature dependence on scales $\ll t$.

The more complex case of QPI for multiple, mutually-interacting magnetic impurities, remains to be investigated. Such systems will display an even wider array of impurity physics, due to the competition of local and non-local (RKKY-type) interactions between impurities. These are expected to have a significant impact on the QPI for randomly distributed impurities, and will be the subject of future work.\cite{multi_imp}


\acknowledgments 
We thank M.~R.~Galpin and R.~Bulla for fruitful discussions. This research was supported by EPSRC grant EP/I032487/1 (AKM,DEL) and the D-ITP consortium, a program of the Netherlands Organisation for Scientific Research (AKM).


\appendix
\section*{Appendix: Convolution method for lattice Green functions}
\label{appx:convmethod}
The calculation of real space lattice Green functions (LGFs) for periodic tight-binding (TB) models is a well-known problem relevant to many areas of physics.\cite{velev,10:Guttmann,00:Cserti} The (retarded) real-space Green function between sites at $\boldr$ and $\boldr'$  on a $d$-dimensional lattice is generally given by,
\begin{align}\label{eq:LGFintRS}
G^0(\boldr,\boldr',\omega)=\int\displaylimits_{1BZ}\frac{d^d\mathbf{k}}{\VBZ}\quad\frac{\E^{-\I(\boldr'-\boldr).\boldk}}{\omega+\I0^+-\epsilon_0-\epsk},
\end{align}
where $\epsk$ is the dispersion and $\epsilon_0\equiv \mu$ is a constant onsite energy (or chemical potential). For a nearest neighbour (NN) TB lattice specified by Eq.~\ref{eq:Hhost}, $\epsk =t\sum_{n}\E^{\I\boldsymbol\delta_n.\boldk}$, with $\{\boldsymbol\delta_n\}$ the set of NN lattice vectors. 

In 1d, simple expressions for the local (on-site) and non-local (inter-site) LGFs are readily obtained, either by direct evaluation of \eref{eq:LGFintRS}, or via equations of motion.\cite{eom,11:MitchellBullaRS}. For the terminal site of a semi-infinite 1d chain with on-site energies $\epsilon_0$, the local LGF is given exactly in closed form by,
\begin{align}
\label{eq:Gs1d}
\begin{split}
G_{\text{1d}}^{0}(\epsilon_0,\omega) = f\left( \frac{\omega-\epsilon_0}{2t}\right ) 
\qquad \text{where}\\
tf(\tilde{\omega}) = \tilde{\omega} -\begin{cases} \sgn{\tilde{\omega}}\sqrt{{\tilde{\omega}}^2-1}\quad &|\tilde{\omega}|>1\\\I\sqrt{1-\tilde{\omega}^2} \quad &|\tilde{\omega}|\leq1\end{cases} 
\end{split} 
\end{align}
which is equivalent to Eq.~\ref{eq:G01d}. The LGFs for the \emph{infinite} 1d chain can be obtained in terms of Eq.~\ref{eq:Gs1d} by exploiting translational invariance,
\begin{align}
\label{eq:1Dinf_LGFdef}
G^{0}_{1\mathrm{d}_\infty}(x,x',\epsilon_0,\omega)=\frac{\left(tG_{\text{1d}}^{0}(\epsilon_0,\omega)\right )^{|x-x'|}}{\omega-\epsilon_0-2t^2G_{\text{1d}}^{0}(\epsilon_0,\omega)} \;.
\end{align}

However, LGFs for various lattice geometries in two and three dimensions are typically highly complicated,\cite{71:KatsuraMoritaINTRO,71:MoritaSC,*71:MoritaRECUR} and not available in closed form. Direct numerical evaluation of \eref{eq:LGFintRS} is notoriously difficult, particularly for low energies, large site separations, or in the vacinity of Van Hove singularities. Recursion relations have been established in several cases, but solutions are often numerically unstable.\cite{09:Berciu,*10:BerciuCook} Improved variants of the recursion technique (or continued fraction expansions) have been developed,\cite{72:HaydockKelly,*75:HaydockKelly,*85:PettiforWeaire,*75:FalicovYndurain,09:Berciu,*10:BerciuCook} but are costly to implement if LGFs are needed as an entire function of frequency.

Here we derive a novel approach to the calculation of LGFs on hypercubic-type lattices, which is both highly accurate and numerically efficient. The method exploits the simple closed form expressions for the LGFs in 1d, Eqs.~\ref{eq:Gs1d} and \ref{eq:1Dinf_LGFdef}, building up lattices in higher dimensions by successive convolutions of those functions. The process is highly efficient because fast Fourier transform algorithms can be used to perform the convolution integrals. The method also has the advantage that boundary edges in 2d systems or explicit surfaces in 3d systems can be simply treated. Indeed, 2d nanoribbon or 3d block geometries can be implemented; and infinite systems can also be handled directly with no extra cost. 

We demonstrate the method first for the infinite 2d square lattice (and henceforth set $\epsilon_0=0$ without loss of generality). We denote the creation operator at lattice site $\boldr=(x,y)$ as $\cre{c}{\boldr,\sigma}\equiv\creIndex{c}{y}{x}$, and define the vector of operators for row $y$ as $\veccre{c}{y,\sigma}=(... \,, \creIndex{c}{y}{1},\, \creIndex{c}{y}{2},\, \creIndex{c}{y}{3},\, ...)$. The TB Hamiltonian then takes the form,
\begin{align}
\label{eq:2D_TB}
\ham_{2d}=\sum_{y=-\infty}^{\infty}\sum_{\sigma}\left [ \veccre{c}{y,\sigma}\hat{T}_{1d}\vecann{c}{y,\sigma} - t\left(\veccre{c}{y,\sigma}\vecann{c}{y+1,\sigma}+ \mathrm{H.c.}\right) \right ] \;,
\end{align}
where $\hat{T}_{1d}$ a matrix describing the connectivity between sites of the $y$-th row (here equal to $t$ for nearest-neighbor sites and 0 otherwise). Importantly, for hypercubic-type lattices, $\hat{T}_{1d}$ is independent of row index $y$. Eq.~\ref{eq:2D_TB} represents a set of \emph{coupled} infinite 1d TB chains to form the 2d lattice. 

We now perform a canonical transformation of operators $\vecann{f}{y,\sigma}=\hat{U}^{\dagger}\vecann{c}{y,\sigma}$, with the matrix $\hat{U}$ defined such that $\hat{D}=\hat{U}^{\dagger}\hat{T}_{1d}\hat{U}$ is diagonal ($D_{kk'}=\epsilon^{1d}_k \delta_{kk'}$). Since this system is infinite and periodic in the $x$-direction, $k\equiv k_x$ can be understood as the Bloch momentum, and $\epsilon^{1d}_k$ the 1d dispersion. However, in general (e.g. for systems with a boundary), $k$ merely labels an eigenstate of $\hat{U}$, with eigenvalue $\epsilon^{1d}_k$.

\begin{figure}
   \subfloat[\ 2d square lattice]{\includegraphics[width=8cm]{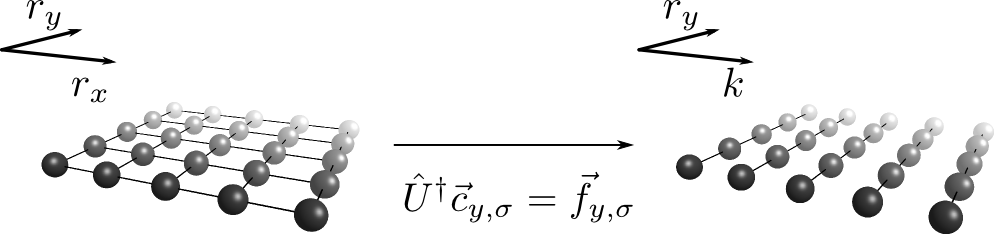}\label{appx_fig:2dTrans}}\\
   \subfloat[\ 3d cubic lattice with (100) surface]{\includegraphics[width=8cm]{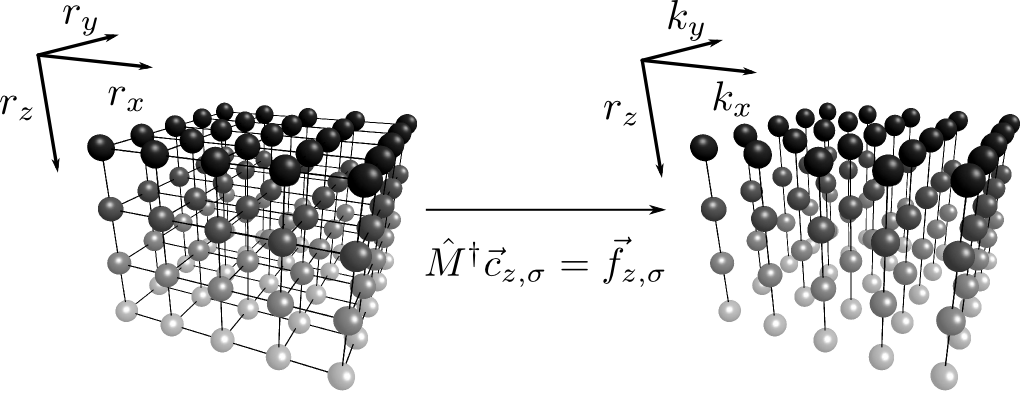}\label{appx_fig:3dsurfTrans}}
\caption{\label{appx_fig:chainTrans} Schematic showing the transformation from the real-space basis (left) to a basis of \emph{decoupled} 1d chains (right). Bonds denote hoppings {$t$} connecting sites. {$\hat{U}$} diagonalizes rows of constant $r_y$ in the infinite 2d square lattice system of (a); while $\hat{M}$ diagonalizes planes of constant $r_z\ge 0$ in the semi-infinite 3d cubic lattice system of (b).}
\end{figure}

In this basis, \eref{eq:2D_TB} reduces to
\begin{align}
\label{eq:2D_decoupledChains}
\ham_{2d}=\sum_{k}\left[\sum_{y,\sigma}\epsilon_{k}\creIndex{f}{y}{k}\annIndex{f}{y}{k}-t\left(\creIndex{f}{y}{k}\annIndex{f}{y+1}{k}+\mathrm{H.c.}\right)\right] \;,
\end{align}
which describes a set of \emph{decoupled} 1d chains labelled by $k$, each with constant on-site energy $\epsilon^{1d}_k$. The transformation from coupled to uncoupled chains is 
shown schematically in \fref{appx_fig:2dTrans}.

The LGFs can then be expressed as,
\begin{align}
\begin{aligned}\label{eq:2D_LGFdef}
G_{2d}^0(\boldr,\boldr',\omega)&\equiv\llangle\annIndex{c}{y}{x};\creIndex{c}{y'}{x'}\rrangle_\omega \\
&=\sum_{k,k'}U^{\phantom{*}}_{xk}U^{*}_{k'x'}\llangle\annIndex{f}{y}{k};\creIndex{f}{y'}{k'}\rrangle_\omega \\
&=\sum_{k}U^{\phantom{*}}_{xk}U^{*}_{kx'} G^{0}_{1\mathrm{d}_\infty}(y,y',\epsilon^{1d}_k,\omega) \;,
\end{aligned}
\end{align}
where the last line follows because $\llangle\annIndex{f}{y}{k};\creIndex{f}{y'}{k'}\rrangle_\omega\propto \delta_{kk'}$ is diagonal in $k$ (see \fref{appx_fig:2dTrans}). The 1d Green function $G^{0}_{1\mathrm{d}_\infty}(y,y',\epsilon^{1d}_k,\omega)$ is given by Eq.~\ref{eq:1Dinf_LGFdef}. 

We now make use of the spectral representation of the 1d LFGs, which can be expressed in terms of $\epsilon^{1d}_k$ and $U_{xk}$, viz:
\begin{align}
\mathrm{Im}\,G^{0}_{1\mathrm{d}_\infty}(x,x',0,\omega')= -\pi\sum_k U^{\phantom{*}}_{xk}U^{*}_{kx'}\delta(\omega'-\epsilon^{1d}_k) \;.
\end{align}
One can then write \eref{eq:2D_LGFdef} as,
\begin{align}
\begin{aligned}\label{eq:2D_LGFconv}
G_{2d}^0(\boldr,\boldr',\omega)=-\frac{1}{\pi}\int_{-\infty}^{\infty}d\omega'~\mathrm{Im}~&G^{0}_{1\mathrm{d}_\infty}(x,x',0,\omega') \\ \times &G^{0}_{1\mathrm{d}_\infty}(y,y',\omega',\omega) \;.
\end{aligned}
\end{align}
Since $G^{0}_{1\mathrm{d}_\infty}(y,y',\omega',\omega)\equiv G^{0}_{1\mathrm{d}_\infty}(y-y',\omega-\omega')$ from Eqs.~\ref{eq:Gs1d} and \ref{eq:1Dinf_LGFdef}, Eq.~\ref{eq:2D_LGFconv} takes the form of a convolution integral. Convolution theorem then allows the efficiency of fast Fourier transform algorithms to be exploited, as
\begin{align}\begin{aligned}
\mathcal{F}_\omega[G^0(\boldr,\boldr',\omega)]=-\tfrac{1}{\pi}\mathcal{F}_\omega[\mathrm{Im}\,&G^{0}_{1\mathrm{d}_\infty}(x,x',\omega)]\\&\times\mathcal{F}_\omega[G^{0}_{1\mathrm{d}_\infty}(y,y',\omega)],
\end{aligned}\end{align}
where $\mathcal{F}_\omega$ denotes Fourier transformation.

Straightforward extension of this method allows access to LGFs in higher dimensions. As a final instructive example, we consider now the 3d cubic lattice with an explicit (100) surface. The Hamiltonian is written as,
\begin{align}
\ham_{3d}=\sum_{z=0}^{\infty}\sum_{\sigma} \left [ \veccre{c}{z,\sigma}\hat{T}_{2d} \vecann{c}{z,\sigma}-t\left(\veccre{c}{z,\sigma}\vecann{c}{z+1,\sigma}+\mathrm{H.c.}\right) \right ]\;,
\end{align}
in terms of vectors of operators for \emph{planes} stacked in the $z$-direction, $\veccre{c}{z,\sigma}=(... \,, \creIndex{c}{z}{\boldr_1},\, \creIndex{c}{z}{\boldr_2},\, \creIndex{c}{z}{\boldr_3},\, ...)$, where $\creIndex{c}{z}{\boldr}$ creates an electron at site $\boldr=(x,y)$ of plane $z$. $\hat{T}_{2d}$ is now the connectivity matrix for the 2d square lattice planes.

We now diagonalize each 2d plane by writing $\vecann{f}{z,\sigma}=\hat{M}^{\dagger}\vecann{c}{z,\sigma}$ such that $\hat{D}=\hat{M}^{\dagger}\hat{T}_{2d}\hat{M}$ is diagonal. As before, $D_{kk'}=\epsilon^{2d}_k \delta_{kk'}$, but now $\epsilon^{2d}_k$ is the 2d square lattice dispersion. In the transformed basis, the semi-infinite 3d cubic lattice becomes a bundle of decoupled semi-infinite 1d chains, each with on-site energy $\epsilon^{2d}_k$, as depicted in \fref{appx_fig:3dsurfTrans}
\begin{align}\label{eq:3D_decoupledChains}
\ham_{3d}=\sum_{k,\sigma}\left [\sum_{z=0}^{\infty}\epsilon^{2d}_k\creIndex{f}{z}{k}\annIndex{f}{z}{k}-t\left(\creIndex{f}{z}{k}\annIndex{f}{z+1}{k}+\mathrm{H.c.}\right)\right ] \;.
\end{align}

The surface LGFs, with $z=0$, then follow as
\begin{align}\label{eq:3D_LGFdef}
\begin{aligned}
G_{\text{surf}}^0(\boldr,\boldr',\omega)\equiv&\llangle \annIndex{c}{0}{\boldr};\creIndex{c}{0}{\boldr'}\rrangle_\omega \\ =& \sum_{k} M^{\phantom{*}}_{\boldr k} M^*_{k \boldr'} G^{0}_{1\mathrm{d}}(\epsilon^{2d}_k,\omega) \;,
\end{aligned}
\end{align}
where $G^{0}_{1\mathrm{d}}(\epsilon^{2d}_k,\omega)$ is given by Eq.~\ref{eq:Gs1d}. Employing the spectral representation of the 2d square lattice Green functions,
\begin{align}
\mathrm{Im}\,G_{2d}^{0}(\boldr,\boldr',\omega)= -\pi\sum_k M^{\phantom{*}}_{\boldr k}M^{*}_{k \boldr'}\delta(\omega-\epsilon^{2d}_k),
\end{align}
we can write
\begin{align}\begin{aligned}\label{eq:3D_LGFconv}
G_{\text{surf}}^0(\boldr,\boldr',\omega)=-\frac{1}{\pi}\int_{-\infty}^{\infty}d\omega'~\mathrm{Im}~&G_{2d}^{0}(\boldr,\boldr',\omega') \\ \times &G^{0}_{1d}(\omega',\omega)\bigg\},
\end{aligned}\end{align}
which can again be viewed as a convolution, here between the semi-infinite 1d Green function $G^{0}_{1d}(\omega',\omega)\equiv G^{0}_{1d}(\omega-\omega')$ given in Eq.~\ref{eq:Gs1d}, and the 2d square lattice Green function $G_{2d}^{0}(\boldr,\boldr',\omega)$ given in Eq.~\ref{eq:2D_LGFconv}.


 %


\begin{thebibliography}{83}
\bibitem [{not({\natexlab{a}})}]{note:eta}%
  \BibitemOpen
  \bibinfo {note} {For numerical calculations, poles are broadened using
  complex frequency $\omega+\I\eta$ with finite $\eta$. For the QPI color
  plots, we use $\eta/t=10^{-3}$, while for Brillouin zone cuts we use
  $\eta/t=10^{-4}$ for improved accuracy near divergences.}\BibitemShut {Stop}%
\end{thebibliography}
\end{document}